\documentclass[10pt,journal,compsoc]{IEEEtran}

\usepackage{setspace} 
\usepackage{latexsym, graphicx, textcomp}
\usepackage{epsfig,upgreek}
\usepackage{amsfonts, amssymb, amsthm,makecell,enumerate}
\usepackage{float,color,dblfloatfix}
\usepackage{fancyhdr}
\usepackage{amsmath}
\usepackage{amsfonts}
\usepackage{mathrsfs}
\usepackage{pifont}
\usepackage{amssymb}
\usepackage{graphicx}
\usepackage{multicol}
\usepackage{cases}
\usepackage{epstopdf}
\usepackage{ltxtable, filecontents}
\usepackage{latexsym}
\usepackage{booktabs}
\usepackage{multirow}
\usepackage{booktabs}
\usepackage{threeparttable}
\usepackage{supertabular}
\usepackage[figuresright]{rotating}
\usepackage[tight,small]{subfigure}
\usepackage{tabularx}
\usepackage{cite}
\usepackage{dcolumn}
\usepackage{algorithmicx}
\usepackage[linesnumbered,ruled,vlined]{algorithm2e}
\usepackage{algpseudocode}
 \usepackage{ragged2e}
 \usepackage{cuted}
\usepackage{graphicx} 
\usepackage{diagbox}
\usepackage{array}
\usepackage{bm}

\graphicspath{{figures/}}

\begin{document}

\title{Energy-Aware Graph Task Scheduling in Software-Defined Air-Ground Integrated Vehicular Networks}

\author{Minghui Liwang, \IEEEmembership{Member}, \IEEEmembership{IEEE}, Zhibin Gao, \IEEEmembership{Member}, \IEEEmembership{IEEE}, Xianbin Wang,~\IEEEmembership{Fellow}, \IEEEmembership{IEEE}

\thanks{Minghui Liwang (minghuilw@xmu.edu.cn) and Zhibin Gao (gaozhibin@xmu.edu.cn) are with the Department of Information and Communication Engineering, School of Informatics, Xiamen University, Fujian, China. Xianbin Wang (xianbin.wang@uwo.ca) is with the Department of Electrical and Computer Engineering, Western University, Ontario, Canada. }}


\IEEEtitleabstractindextext{
\begin{abstract}
\justifying
The \underline{S}oftware \underline{D}efined \underline{A}ir-\underline{G}round integrated \underline{V}ehicular (SD-AGV) networks have emerged as a promising paradigm, which realize the flexible on-ground resource sharing to support innovative applications for UAVs with  heavy computational overhead. In this paper, we investigate a vehicular cloud-assisted task scheduling problem in SD-AGV networks, where the computation-intensive tasks carried by UAVs, and the vehicular cloud are modeled via graph-based representation. To map each component of the graph tasks to a feasible vehicle, while achieving the trade-off among minimizing UAVs' task completion time, energy consumption, and the data exchange cost among moving vehicles, we formulate the problem as a mixed-integer non-linear programming problem, which is Np-hard. Moreover, the constraint associated with preserving task structures poses addressing the subgraph isomorphism problem over dynamic vehicular topology, that further complicates the algorithm design. Motivated by which, we propose an efficient decoupled approach by separating the template (feasible mappings between components and vehicles) searching from the transmission power allocation. For the former, we present an efficient algorithm of searching for all the isomorphic subgraphs with low computation complexity. For the latter, we introduce a power allocation algorithm by applying $p$-norm and convex optimization techniques. Extensive simulations demonstrate that the proposed approach outperforms the benchmark methods considering various problem sizes.

\end{abstract}

}

\maketitle


\IEEEpeerreviewmaketitle

\section{Introduction}

\noindent
\IEEEPARstart{B}{enefiting} from the considerable support in military, public and civil services~\cite{1}, unmanned aerial vehicles (UAVs, also known as drones)~\cite{2} are emerged as one of the fastest growing techniques owing to their freedom of movement, on-demand deployment, and hardware cost reduction~\cite{3}. Last five years have witnessed an exponential growth of UAV applications, presently driving business close to 1 billion US Dollars in the USA, while with an upwards growth targeted to reach 46 billion
US Dollars by 2025~\cite{4}. Specifically, innovative applications such as transportation management, disaster relief (e.g., rescue missions and target detection) and smart surveillance (e.g., tracking mobile targets, etc.) have facilitated both safety and convenience to people's work and life. Besides, UAVs have realized significant values in popular incidents such like the radiation leakages of the Fukushima nuclear power plant in 2011~\cite{5}, and the search work during the basketball legend Kobe Bryant's helicopter crush, in 2020~\cite{6}. 


Computational overhead required by the aforementioned UAV applications and use cases pose major challenges to UAVs with limited on-board processing capabilities, resources, and battery lives (e.g., up to 90 minutes flight duration even for high-performance battery-powered multirotor UAVs)
~\cite{7,8}, which calls for responsive and flexible computing services.
Thus, promoted by extensive development efforts on connected smart cars, vehicular cloud computing (VCC) technology~\cite{9,10,11} facilitates a revolution of the resource sharing among on-ground vehicles with surplus resources, and the UAVs with heavy workloads. Specifically, vehicles (service providers, SPs) form a cloud (vehicular cloud, VC) via vehicle-to-vehicle (V2V) communications to support efficient collaborative computing, while UAVs (service requestors) are encouraged to offload application data to SPs through air-to-ground (A2G) links~\cite{12,13,14}.

Nevertheless, traditional network architecture can hardly meet different QoS requirements imposed by diverse services and various communication techniques in networks containing both air segment (UAVs) and ground segment (vehicles)~\cite{15,16}. To enable a dynamic and adaptable air-ground integrated network with cost-effectivity, software defined networking (SDN) has been applied as an emerging architecture. Specifically, SDN separates the control plane and the data plane, introduces a logically centralized control with a global view of the network, while facilitating network programmability/reconfiguration through open interfaces~\cite{12,16,17}. Motivated by which, we consider the software defined air-ground integrated vehicular (SD-AGV) network architecture, to implement agile and flexible support for heterogenous applications. 
Under such a framework, graph-based-representation~\cite{9,10,18,8,19} is utilized in this paper to characterize the non-negligible internal structures associated with the applications of UAVs. Each application\footnote{The ``application'' is interchangeable with ``task'' or ``graph task'' for the rest of this paper.} is modeled as an undirected graph, where the vertices (named as ``components'' in this paper) represent either data sources or data processing units, while weighted edges describe the dependency among the vertices. Specifically, components can be processed on different SPs in parallel while intermediate data exchanges may happen among these vehicles (associated edges in each task). Moreover, virtual machine (VM)-based representation is utilized to quantify available resources on SPs. 

\subsection{An Example of unDirected Graph Task and the Relevant Scheduling} 
Fig. 1 shows an example of multi-target recognition application in smart public transportation which is popular nowadays~\cite{8} (e.g., masked face recognition smart buses), especially during the COVID-19 epidemic~\cite{20} to trace the close contacts of patients who are tested positive. Specifically, a task is modeled as an undirected graph where data of multiple targets, e.g., faces, can be analyzed on different servers in parallel. Meanwhile, edges represent the interdependency among components such as outline and color information, etc (e.g., to indicate close contacts during epidemic). Thus, each edge in the graph task may require intermediate data exchange between the two SPs who are handling components associated with this edge. Similar paradigm also refers to image segmentation-based parallel processing, VR video stream processing~\cite{8}, as well as federated learning-based applications (e.g., clients share a global model and train their own models via local data sets in parallel)\footnote{Note that the most common bit-stream-based computation-intensive application~\cite{13,23,24} can also be seen as a specific graph task with one divisible component, with no edges.}. 

\begin{figure}[h!t]
\centering
\includegraphics[width=.95\linewidth]{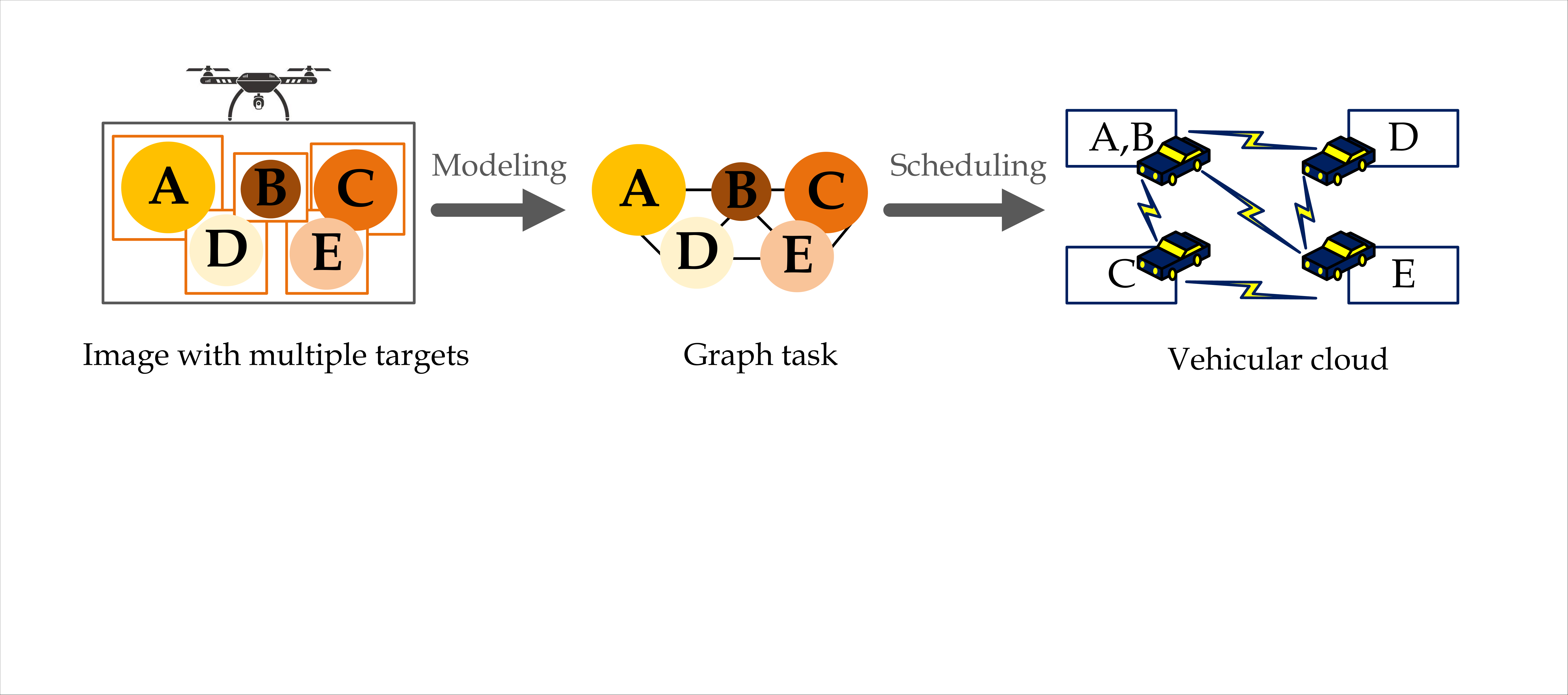}
\caption{An toy example of graph task under un-directed graph representation.}
\end{figure}

\subsection{Challenges} 
In this paper, we study an interesting energy-aware graph task scheduling problem in a SD-AGV network architecture. Concretely, the components of graph tasks carried by UAVs can be mapped to feasible on-ground SPs, while achieving the trade-off upon minimizing the task completion time and the energy consumption of UAVs, as well as the data exchange cost (e.g., energy) among SPs which incurred by the required data interactions among task components. Major challenges are summarized below, 

\noindent
\textit{a}) Obtaining feasible mappings between the components of graph tasks and the moving SPs requires solving the \textit{subgraph isomorphism} problem under contact constraints, which is NP-hard~\cite{21,22}. Moreover, avoiding the conflict of premium resources (e.g., some vehicles are at the core of vehicular cloud topology) can further pose challenges.


\noindent
\textit{b}) Considering UAVs' energy consumption requires addressing the transmission power optimization problem, which is generally formulated with \textit{non-convex} feature;

\noindent
\textit{c}) The energy-aware graph task scheduling problem stands for a coupling of obtaining the mappings between the components and the SPs, as well as solving the transmission power optimization problem, which are challenging to be solved in parallel.

Addressing the above-mentioned challenges represents our main motivation in this paper. Specifically, we investigate a novel decoupled approach to solve the energy-aware graph task scheduling problem, by separating components mapping from power allocation. For the former, an efficient template search algorithm is proposed, where each template stands for a feasible mapping between the components of graph tasks and the SPs. For the latter, a power allocation algorithm is introduced via applying $p$-norm and convex optimization techniques.

\subsection{Related Work}


\noindent
Existing studies devoted to the task scheduling/allocation problem can be roughly divided into two categories based on task models: \textit{a}) tasks that are represented by bit streams without concerning the inherent dependencies, such as~\cite{1,13,23,24,25,26,27}; \textit{b}) tasks under graph-based representation upon considering the inner dependencies among components such like~\cite{9,10,28,29,18,30,31,32,33,34,22,19}, which stands for our main focus. The bit stream-represented task scheduling problem has been widely discussed in UAV networks. Messous \textit{et al.} in~\cite{1} focused on the computation offloading problem in a mobile edge computing (MEC)-assisted UAV network, by establishing a non-cooperative theoretical game with multi-players and three pure strategies. In~\cite{13}, Wang \textit{et al.} introduced a vehicular fog computing (VFC) system where vehicles perform computation tasks offloaded from UAVs. Specifically, a stable matching algorithm was proposed to avoid the transmission competitions yet enabling cooperations among UAVs and vehicles.
Bai \textit{et al}.~\cite{23} conceived an energy-efficient computation offloading technique for UAV-MEC systems, with an emphasis on physical-layer security. 
Sacco \textit{et al}.~\cite{24} addressed the problem of task offloading from UAV to the closest edge cloud via introducing a simple yet effective formalization that enables a learning process, while reducing the required information and training time.

\begin{figure*}[t!]
\centering
\subfigure[]{\includegraphics[width=.467\linewidth]{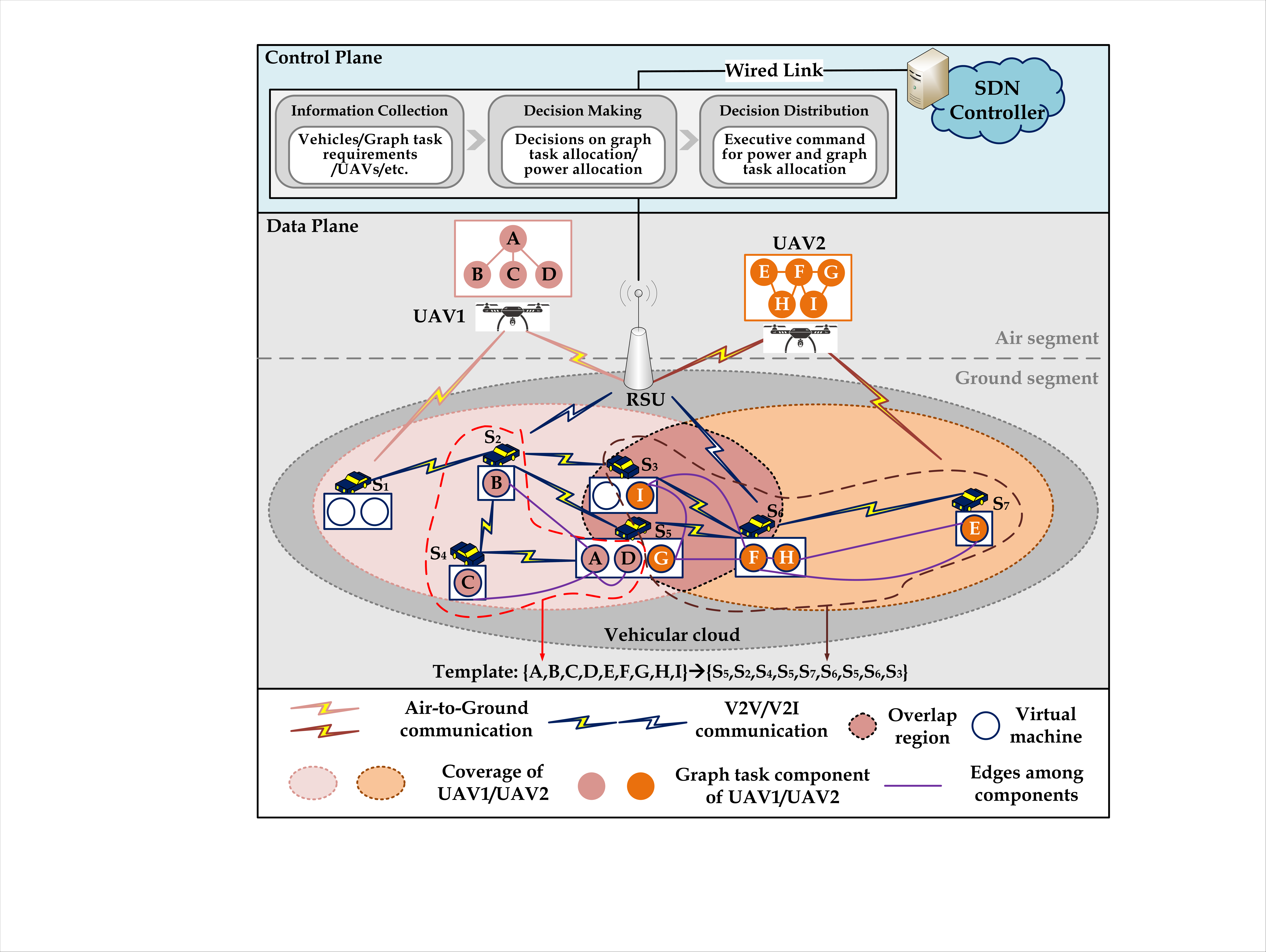}}~~~~~
\subfigure[]{\includegraphics[width=.484\linewidth]{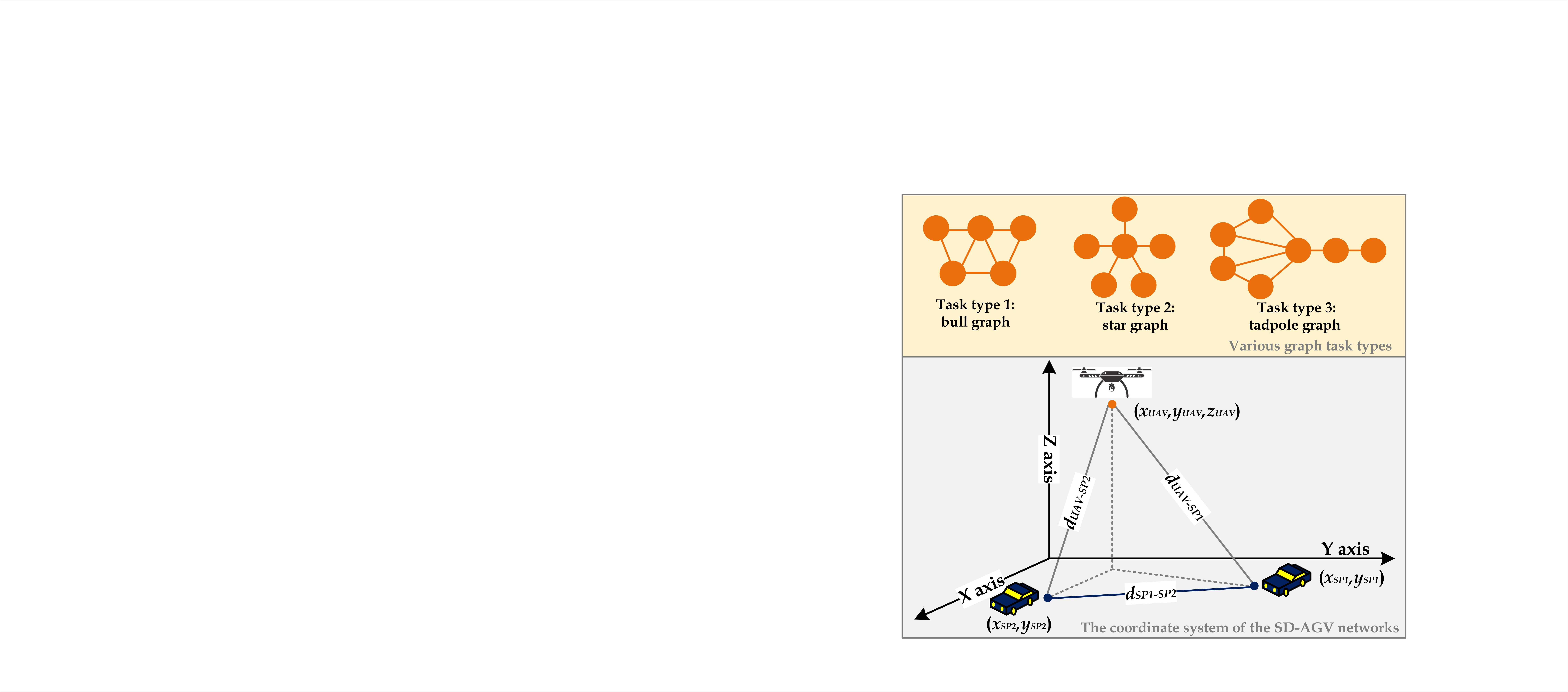}}
\caption{a). The framework of VC-assisted graph task scheduling in the SD-AGV network; b). The related coordinate system, and various graph task types considered in simulation.}
\end{figure*}

Studies associated with graph task scheduling can be further classified into two types according to the dependency among task components: \textit{a}) directed acyclic graph (DAG) task scheduling~\cite{30,32,33,34,19}; and \textit{b}) undirected graph-represented task scheduling\cite{9,10,28,29,18,31,22}. 
Regarding DAG-based task models, Sahni \textit{et al.} in \cite{30} formulated the problem of jointly offloading multiple DAG tasks and network flow (considering start time of network flow) scheduling in collaborative edge computing to minimize the average task completion time. Goudarzi \textit{et al}.~\cite{32} proposed a fast hybrid multi-site computation offloading algorithm by modeling each application as a weighted related graph. 
Geng \textit{et al}. addressed the energy-efficient computation offloading problem on multicore-based mobile devices in~\cite{33} by formulating a mixed-integer nonlinear programming problem and applying a heuristic algorithm. In~\cite{34}, Sun \textit{et al}. studied a VC-based computation offloading mechanism where computing missions were modeled as tasks with interdependency and executed in different vehicles to minimize overall response time, and thus alleviates the heavy workloads of edge clouds. Considering static data center networks, Shafiee \textit{et al}. in \cite{19} proposed a polynomial-time algorithm while proving an optimality gap for scheduling coflows with general DAGs. 

For un-directed graph models, considering static topologies of both computing servers and users in cloud computing context, Ghaderi \textit{et al}.~\cite{18} proposed a randomized graph job scheduling algorithm by considering job arrivals/departures, which facilitated a smooth trade-off between the average execution cost and queue length. 
An energy-efficient graph job allocation framework in geo-distributed cloud networks was proposed by Hosseinalipour \textit{et al}.~\cite{28}, where solutions were obtained for data center networks of various scales. 
In~\cite{31}, aiming to minimize the job completion time while considering energy consumption, the problem of scheduling embarrassingly parallel jobs composed of a set of independent tasks was studied by Shi \textit{et al}. In general, parallel processing of components of un-directed graph tasks greatly require servers to maintain all the communication links in supporting intermediate data sharing, which is fundamentally different from the processing of DAG models \cite{33,19} that are sequential in nature. Besides, considering mobile devices as computing servers can further pose challenges to protect the structure of graph tasks over dynamic service topology. 


We are among the few works that study the undirected graph task scheduling problem considering dynamic service providers (e.g., vehicles). A randomized graph task allocation mechanism based on hierarchical tree decomposition was proposed in our previous work~\cite{9}, through which, feasible mappings between components and SPs were obtained. In~\cite{10}, we presented a novel multi-task offloading problem under graph-representation by considering the potential inter-component competition due to task concurrency. In \cite{22}, we studied a truthful auction-based graph task allocation problem in vehicular cloud-assisted networks while considering the resource reutilization. However, concerning energy consumption especially for battery-conscious devices are not yet considered in the above-mentioned previous researches. To conform green computing~\cite{35}, an energy-aware graph task allocation problem in VC-assisted IoV was studied in our latest study~\cite{29}. Specifically, a hierarchical tree based random matching approach was applied to determine the assignment of a single graph task over service topology; while a structure-preserved simulated annealing algorithm was proposed to solve the power allocation problem.
\subsection{Contributions}

\noindent
This paper studies the energy-aware allocation problem of mapping the components of multiple graph tasks carried by UAVs to on-ground SPs. Factors such as multiple concurrent tasks and complicated A2G channels present additional challenges related to the problem size and algorithm efficiency. Moreover, the potential competitions among UAVs caused by communication overlaps are necessarily considered. To the best of our knowledge, this paper is among the first to study the energy-aware graph task scheduling problem under the SD-AGV network architecture. Major contributions are summarized below:

\noindent
$\bullet$ Framework of energy-aware graph task scheduling under SD-AGV network is introduced, where the SDN controller achieves the efficient orchestration of undeveloped on-ground resources. The reliable integration of air segment and ground segment enables resource sharing between the UAVs with computation-intensive graph tasks and the vehicles with idle resources, under the logically centralized control of the SDN controller.

\noindent
$\bullet$ An interesting graph task scheduling problem under SD-AGV network architecture is studied, where each task is modeled as an undirected graph with components and weighted edges. Besides, VC (topology of SPs) is modeled as a service graph where probabilistic representation is utilized to evaluate the communications among SPs. Through solving the problem, the components of graph tasks on UAVs can be efficiently mapped to feasible on-ground vehicles, while achieving the trade-off upon minimizing the task completion time and energy consumption of UAVs, as well as the data exchange cost (e.g., energy consumption of SPs) among SPs.

\noindent
$\bullet$ The afore-mentioned problem is formulated as a mixed integer non-linear programming (MINLP) problem, which is NP-hard. Moreover, one of the constraints associated with preserving the graph task structures requires addressing the subgraph isomorphism problem, which further complicates the algorithm design. Thus, we propose an ingenious decoupled approach by separating the template search stage from the power allocation stage. The problem in the former stage is formulated as searching for all the isomorphic subgraphs between the graph tasks and the VC, for which we present an efficient template search algorithm. For the latter stage, we introduce a practical power optimization algorithm by applying convex optimization techniques.

\noindent
$\bullet$ Based on thorough numerical analysis and comparative evaluations, we demonstrate that the performance of the proposed decoupled approach can outperform the baseline methods considering various problem sizes, while providing a low computation complexity in most cases.

The rest of this paper is organized as follows. The system model is introduced in Section~2. We formulate the energy-aware graph task scheduling problem as a MINLP problem in Section~3. In Section~4, we propose an efficient decoupled approach. The performance evaluation through comprehensive simulations is introduced in Section~5 before drawing the conclusion in Section~6.

\section{System Model}

\begin{table*}[t!]
{\small
\center
\caption{Major notations and explanations}
\begin{tabular}{ll}
\hline\\[-2.9mm]\hline
Notation & Explanation \\
\hline
${\bm{G}}^{\bm{s}}, \bm{S}, {\bm{E}}^{\bm{s}}, {\bm{W}}^{\bm{s}}$ & the VC graph, the set of SPs, edges and weights \\
$s_k$, $e^s_{k, k'}$, $ w^s_{k, k'}$ & the $k^{\mathrm{th}}$ SP, the edge and weight between SPs $s_k$ and $s_{k'}$ \\
$\bm{U}$, $u_m$, $\bm{{\mathcal{R}}_{m}}$ & the set of UAVs, the $m^{\mathrm{th}}$ UAV, the SPs set covered by $u_m$ \\
${\bm{G}}^{{\bm{u}}_{\bm{m}}}$, ${\bm{V}}^{{\bm{u}}_{\bm{m}}}, {\bm{E}}^{{\bm{u}}_{\bm{m}}}, {\bm{W}}^{{\bm{u}}_{\bm{m}}}$ & the graph task of $u_m$, the set of components, edges, and weights of ${\bm{G}}^{{\bm{u}}_{\bm{m}}}$ \\
$v_{n, m}$, $ e^{u_m}_{n, n'}$, $ w^{u_m}_{n, n'}$ & the $n^{\mathrm{th}}$ components in ${\bm{V}}^{{\bm{u}}_{\bm{m}}}$, the edge and weight between components $v_{n, m}$ and $v_{n', m}$ \\
$x^{n, m}_k$, $ q_{k, m}$ & the indicator of mapping $v_{n, m}$ to $s_k$, the allocated power of $u_m$ to $s_k$ \\
$\bm{x}, \bm{q}$ & the matrix of $x^{n, m}_k$ and $q_{k, m}$ \\
$d_{k, k'}$, $d^{A2G}_{k, m}$ & the distance between $s_k$ and $s_{k'}$ the distance between $u_m$ and $s_k$ \\
${\mathcal{C}}^{n, n', m}_{k, k'}$ & indicator of the data exchange cost \\
$g_{k, m}$, $ r_{k, m}$ & the channel gain and the data transmission rate between $u_m$ and $s_k$ \\
$D_{n, m}$, $ t^{exec}$ & the data size of component $v_{n, m}$, the execution time of each VM \\
${\alpha}_1$, $ {\alpha}_2$ & the probabilistic parameters \\
$\bm{X}$, $\bm{{\mathcal{X}_z}}$ & the template set, the $z^{\mathrm{th}}$ template in set $\bm{X}$ (used in stage 1) \\
$\widetilde{{\bm{S}}}_{\bm{z}}$, $\widetilde{{\bm{E}}}^{\bm{s}, \bm{z}}$, $ \widetilde{{\bm{W}}}^{\bm{s}, \bm{z}}$ & the set of SPs, edges and weights associated with template ${\mathcal{X}}_{\bm{z}}$ (used in stage 1) \\
${\tilde{s}}^z_k$, $ {\tilde{e}}^{s, z}_{k, k'}$, $ {\tilde{w}}^{s, z}_{k, k'}$ & the corresponding SP, edge and weight in sets $\widetilde{{\bm{S}}}_{\bm{z}}$, $\widetilde{{\bm{E}}}^{\bm{s}, \bm{z}}$ and $ \widetilde{{\bm{W}}}^{\bm{s}, \bm{z}}$ (used in stage 1) \\
$\bm{N}$ & the component exploration sequence \\
${\mathcal{D}}^c (v_{n, m})$, $ {\mathcal{D}}^{map} (v_{n, m})$ & the degree, and the component mapping degree of $v_{n, m}$ \\
${\mathcal{D}}^s (s_k)$ & the current available degree of $s_k$ \\
$\bm{Pred}({\bm{v}}_{\bm{n}, \bm{m}})$ & the set of predecessors of $v_{n, m}$ \\
${\bm{V}}^{\bm{u}}, {\bm{E}}^{\bm{u}}$, ${\bm{W}}^{\bm{u}}$ & the set of components, edges and weights regardless of any particular UAV or template (used in stage 2) \\
$v_n$, $ e^u_{n, n'}$, $ w^u_{n, n'}$ & the corresponding component, edge and weight in sets ${\bm{V}}^{\bm{u}}, {\bm{E}}^{\bm{u}}$ and ${\bm{W}}^{\bm{u}}$ (used in stage 2)\\
$\widetilde{\bm{S}}$, $ \widetilde{{\bm{E}}}^{\bm{s}}$, $ \widetilde{{\bm{W}}}^{\bm{s}}$ & the set of SPs, edges and weights regardless of any particular template (used in stage 2)\\
${\tilde{s}}_k$, $ {\tilde{e}}^s_{k, k'}$, $ {\tilde{w}}^s_{k, k'}$ & the corresponding SP, edge and weight in sets $\widetilde{\bm{S}}$, $\widetilde{{\bm{E}}}^{\bm{s}}$ and $\widetilde{{\bm{W}}}^{\bm{s}}$ (used in stage 2) \\
\hline\\[-2.9mm]\hline
\end{tabular}
\label{tab1}
}
\end{table*}

\subsection{The Framework of Graph Task Allocation in SD-AGV Networks} 

\noindent
The SD-AGV networks is seen as an emerging network architecture of the late years. Specifically, SDN\footnote{Notably, this paper does not go into details of the internal mechanisms of SDN, which stands for a background since multiple graph task assignment calls for the overall knowledge of service topology. Existing papers with similar ideas are also supportive, e.g., \cite{30}.} decouples the control plane from the data plane~\cite{12}, introduces logically centralized control with a global view of networks, while facilitating network programmability/reconfiguration through open interfaces. Combine with the VCC technology, a manageable and cost-effective marketplace is established to orchestrate on-ground resources for the UAVs with computing requirements, achieving an efficient collaborative computing system. The framework of VC-assisted graph task scheduling in SD-AGV networks and the relevant coordinate system are shown in {Fig}.~2(a) and {Fig}.~2(b), respectively. 

\noindent
\textbf{Data plane and available resources}: In this framework, each UAV or vehicle serves as an SDN switch that abided by unified scheduling and follows the Openflow protocol commonly used in SDN~\cite{17}. The UAVs are service requestors with heavy workloads, while vehicles serve as service providers with available resources. Both parties are following the schedule of the SDN controller. Decoupling data transmission and processing from control stands for one of the key features in this framework, which enables efficient orchestration of undeveloped resources.

\noindent
\textbf{Control plane and the SDN controller}: This framework effectively facilitates the independency between the physical communication channels of control plane and that of the data plane, where the SDN controller can capture the status information~\cite{15} reported periodically by UAVs and vehicles (e.g., channel state information, locations, computing service requirements, current available resources, etc). Specifically, a feasible energy-aware graph task scheduling decision generated by the SDN controller will be distributed to the related mobile devices. Then, the data of graph tasks will be transmitted from UAVs to on-ground vehicles according to the allocation decision via data plane.

\subsection{Model of Vehicular Cloud}

\noindent
Suppose a VC covers a region containing SP set $\bm{S}=\{s_k|k\in\{1,2,\cdots, |\bm{S}|\}\}$, where each $s_k\in \bm{S}$ owns different number of fully connected idle VMs for leasing, and every VM provides the computational capability related to the execution time $t^{exec}$ for processing one component of a graph task. Notably, an available VM can only process one component at a time. Correspondingly, the VC is represented as a graph ${\bm{G}}^{\bm{s}}=\{\bm{S}, {\bm{E}}^{\bm{s}}, {\bm{W}}^{\bm{s}}\}$, where $\bm{S}$ is the set of SPs and each $s_k\in \bm{S}$ can provide a set of available VMs denoted by $\bm{\mathcal{V}}_{\bm{k}}$. ${\bm{E}}^{\bm{s}} = \{e^s_{k, k'}|s_k, s_{k'}\in \bm{S}, s_k\neq s_{k'}\}$ represents the edge set where $e^s_{k, k'}$ indicates the edge between $s_k$ and $s_{k'}$, namely, a one-hop V2V communication link between these SPs. Moreover, ${\bm{W}}^{\bm{s}} = \{w^s_{k, k'}| s_k, s_{k'}\in \bm{S}, s_k\neq s_{k'}\}$ denotes the associated weight of the edge set describing the corresponding parameters of the exponential distribution of V2V connections, which will be introduced in the following Section 2.5. 

\subsection{Model of UAVs and Graph Tasks}

\noindent
Consider set of UAVs $\bm{U}=\{u_m | m\in\{1,2,\cdots, |\bm{U}|\}\}$ hovered in the sky, where each $u_m\in \bm{U}$ carries a computation-intensive graph task\footnote{Here, each UAV can also carry multiple graph tasks. This case is regarded as a scenario with multiple virtual UAVs with limited transmission powers, where the proposed approach in this paper is feasible to be implemented. 
} requires to be offloaded to on-ground SPs for execution. The task of $u_m$ is modeled as a graph ${\bm{G}}^{{\bm{u}}_{\bm{m}}}=\{{\bm{V}}^{{\bm{u}}_{\bm{m}}}, {\bm{E}}^{{\bm{u}}_{\bm{m}}}, {\bm{W}}^{{\bm{u}}_{\bm{m}}}\}$, where ${\bm{V}}^{{\bm{u}}_{\bm{m}}} = \{v_{n, m}| n\in \{1,2,\cdots, |{\bm{V}}^{{\bm{u}}_{\bm{m}}}|\}\}$ denotes the components set in the graph task, and $v_{n, m}$ indicates the $n^{\rm th}$ component of the graph task of $u_m$, with data size $D_{n, m}=D $ (bit)\footnote{In this paper, we assume that components have the same data size for analytical simplicity. Notably, the proposed approach can also be applied when considering different data sizes of components.}. ${\bm{E}}^{{\bm{u}}_{\bm{m}}} = \{e^{u_m}_{n, n'}| n, n'\in \{1,2,\cdots, |{\bm{V}}^{{\bm{u}}_{\bm{m}}}|\}, n \neq n'\}$ and ${\bm{W}}^{{\bm{u}}_{\bm{m}}} = \{w^{u_m}_{n, n'}| n, n'\in \{1,2,\cdots, |{\bm{V}}^{{\bm{u}}_{\bm{m}}}|\}, n \neq n'\}$ represent the edge set and the related weight set respectively, where $ e^{u_m}_{n, n'}$ and $w^{u_m}_{n, n'}$ denote the required data flow, and connect duration between components $v_{n, m}$ and $v_{n', m}$. A graph task ${\bm{G}}^{{\bm{u}}_{\bm{m}}}$ describes the internal dependencies of how the computation split among the components in ${\bm{V}}^{{\bm{u}}_{\bm{m}}}$. For notational simplicity, let ${\bm{V}}^{\bm{U}}\triangleq \bigcup_{u_m\in \bm{U}}{{\bm{V}}^{{\bm{u}}_{\bm{m}}}}$ be the union set of the components of graph tasks.

\subsection{Templates}
\noindent
Observe that for any graph task, there exist several ways (an exponential large number) in which the task can be distributed over SPs. Note that multiple graph tasks may exist in the network, a \textbf{\textit{template}} $\bm{\mathcal{X}}$ corresponds to a \textbf{\textit{feasible mapping}} from the components set $\bm{V^U}$ to a subset of $\bm{S}$ in the related VC. An example of the template is given in {Fig}.~2(a). Notably, a mapping fails to preserve the structures or meet the weight requirements among edges of the graph tasks cannot be a template.

\subsection{Model of Communication}
\noindent
For analytical simplicity, let binary indicator $x^{n, m}_k=1$ denote the assignment where component $v_{n, m}$ is mapped to SP $s_k$; otherwise, $x^{n, m}_k=0$. Note that different UAVs may have various communication ranges, let $\bm{{\mathcal{R}}_{m}}$ be the set of SPs that are covered by the communication radius of UAV $u_m$, and the components of $\bm{G^{u_m}}$ can only be mapped to the SPs in $\bm{{\mathcal{R}}_{m}}$. Also, we assume that each UAV stays hovering or moves slightly in the sky, which enables the SPs in each set $\bm{{\mathcal{R}}_{m}}$ remaining unchanged during graph task scheduling. 

\noindent
\textbf{V2V channel model:} The contact duration between SPs $s_k$ and $s_{k'}$ obeys an exponential distribution~\cite{9,10,36} with parameter $w^s_{k, k'}$. Thus, the probability of the contact duration $\Delta {\tau}_{k, k'}$ between $s_k$ and $s_{k'}$ exceeding a certain period $\Delta t$ is given by $prob (\Delta {\tau}_{k, k'}\ge \Delta t|w^s_{k, k'})=e^{-\Delta t{\times w}^s_{k, k'}}$, where the larger $prob (\Delta {\tau}_{k, k'}\ge \Delta t|w^s_{k, k'})$ can bring more assurance for the required data exchange duration between different SPs\footnote{Regarding moving vehicles, we consider an soft assurance by using probabilistic representation, which is also close to real-life networks. }.

The path loss of a V2V communication link is considered by following the dual-slope model~\cite{37}, which is defined as a piece-wise function of the distance $d_{k, k'}$ between $s_k$ and $s_{k'}$.
\begin{align} 
\label{eq1} 
pl(d_{k, k'}) \!=\! 
\begin{cases}
{pl}_0 \!+\! 10{\eta}_1 \log_{10}\left(\dfrac{d_{k, k'}}{d_0}\right) \!+\! X_{\sigma} , \\
& \hspace*{-2.2em} \text{if } d_0 \le  d_{k, k'} \le  d_B \\
{pl}_0+10{\eta}_1 \log_{10}\left(\dfrac{d_{k, k'}}{d_0}\right)+\\[2mm]
10{\eta}_2 \log_{10}\left(\dfrac{d_{k, k'}}{d_B}\right)+X_{\sigma} , & \hspace*{-2.2em} \text{if } d_{k, k'}>d_B 
\end{cases}, 
\end{align} 

\noindent
where $d_0$ is the reference distance, ${pl}_0$ is the path loss at $d_0$, $X_{\sigma} $ denotes a zero-mean normally distributed random variable with a standard deviation of $ \sigma $. Notation ${\eta}_1$ and ${\eta}_2$ denote the path loss exponent before and after distance $d_B$, respectively, and $d_B$ indicates the breakpoint distance which is calculated as~\eqref{eq2}.
\begin{align} 
\label{eq2} 
d_B=\frac{4h_t h_r}{\lambda} -\frac{w}{4}, 
\end{align} 

\noindent
where $h_t$ and $h_r$ are the transmitter's and the receiver's height, and $w$ denotes the wavelength. Here, let $h_t=h_r$ owing to the possible intermediate data exchange that makes $s_k$ and $s_{k'}$ into both transmitter and receiver. Rely on the uncertain channel conditions of V2V communication links, the case where two connected components are mapped to different SPs can bring intermediate data exchange cost, which captures the cost incurred from traffic exchange among different SPs in a VC. Correspondingly, $c^{cost}_{k, k'}$ is defined in~\eqref{eq3}.
\begin{align} 
\label{eq3} 
c^{cost}_{k, k'}=
\begin{cases}
f(pl(d_{k, k'})), & {s}_k\neq s_{k'} \\
0, & \text{otherwise} 
\end{cases}, 
\end{align} 

\noindent
where $f()$ is a monotone increasing function. Apparently, a larger value of $pl(d_{k, k'})$ will bring a higher cost for intermediate data exchange among different vehicles. Let $c^{n, n', m}_{k, k'}$ describe the data exchange cost (e.g., energy consumption, etc.) incurred when two connected components $v_{n, m} $ and $v_{n', m}$ of $ {\bm{G}}^{{\bm{u}}_{\bm{m}}}$ are assigned to different SPs, which is given in~\eqref{eq4}.
\begin{align} 
\label{eq4} 
{\mathcal{C}}^{n, n', m}_{k, k'}=
\begin{cases}
c^{cost}_{k, k'}, & \forall e^{u_m}_{n, n'}\in {\bm{E}}^{{\bm{u}}_{\bm{m}}}, \text{if}~x^{n, m}_k\times x^{n', m}_k=1 \\
0, & \textit{otherwise} 
\end{cases}
\end{align} 

\noindent
\textbf{A2G channel model:} As shown in {Fig}.~2(a), we consider several multi-antenna UAVs capable for offloading the components of a graph task to SPs. Moreover, each SP relies on full-duplex techniques\footnote{Each SP can receive task data from UAVs while exchanging the intermediate data results with other SPs~\cite{40}.} and the self-interference is ignored. Denote the channel gain between $u_m$ and $s_k\in \bm{{\mathcal{R}}_{m}}$ as $g_{k, m}$, which is assumed to be dominated by the line of sight (LoS) path~\cite{38}, shown in~\eqref{eq5}.
\begin{align} 
\label{eq5} 
g_{k, m}={g}_1{\times \left(d^{A2G}_{k, m}\right)}^{{-\eta}_3}, 
\end{align} 
\noindent
where ${g}_{1}$ corresponds to the channel gain at the reference distance of 1~meter, $d^{A2G}_{k, m}$ indicates the A2G distance between $u_m$ and $s_k$, and ${\eta}_{3}$ denotes the path loss exponent of the LoS path. The data transmission rate $r_{k, m}$ between UAV $u_m$ and SP $s_k$ is a function of the transmission power $q_{k, m}$ that $u_m$ allocates to SP $s_k$, calculated by~\eqref{eq6}\footnote{Interference during task data transmission procedure (from UAVs to vehicles) is neglected in this paper for analytical simplicity, while technologies such as OFDMA can also support this assumption~\cite{23,30,22,8}.}where $B$ denotes the channel bandwidth and $N_0$ represents the background noise power. 
\begin{align} 
\label{eq6} 
r_{k, m}=B \log_2\left(1+\frac{q_{k, m}\times g_{k, m}}{N_0}\right) 
\end{align} 

\subsection{Model of Computation and Energy Consumption}

\noindent
The completion time $t_m$ of graph task ${\bm{G}}^{{\bm{u}}_{\bm{m}}}$ is composed by the time of data transmission, execution and the resulting feedback. Notably, the delay on resulting feedback from the SP to the UAV can be ignored owing to a much smaller output data size~\cite{9,10}. Apparently, $t_m$ relies on the slowest processed component of graph task ${\bm{G}}^{{\bm{u}}_{\bm{m}}}$.
\begin{align} 
\label{eq7} 
t_m={\max {\left[\frac{\sum^{|{\bm{V}}^{{\bm{u}}_{\bm{m}}}|}_{n=1}{x^{n, m}_k\times D_{n, m}}}{r_{k, m}}\right]}_{1\le k\le |\bm{S}|}} +t^{exec}, 
\end{align} 
where $\sum^{|{\bm{V}}^{{\bm{u}}_{\bm{m}}}|}_{n=1}{x^{n, m}_k\times D_{n, m}}$ and $\frac{\sum^{|{\bm{V}}^{{\bm{u}}_{\bm{m}}}|}_{n=1}{x^{n, m}_k\times D_{n, m}}}{r_{k, m}}$ represent the total amount of data, and the relevant transmission time of components from $u_m$ to $s_k$, respectively. Moreover, the UAV would incur extra overhead in terms of energy when transmitting data to SPs via wireless access. Thus, the energy consumption $c_m$ of $u_m$ is calculated as:
\begin{align} 
\label{eq8} 
c_m=\sum^{|\bm{S}|}_{k=1}{\sum^{|{\bm{V}}^{{\bm{u}}_{\bm{m}}}|}_{n=1}{x^{n, m}_k\left(\frac{q_{k, m}\times D_{n, m}}{r_{k, m}}\right)}}+\ell, 
\end{align} 
where $ \ell $ indicates the tail energy~\cite{39} given that the UAV will hold the channel for a while even after data transmission. We summarize the major notations and the related definitions in \textbf{Table 1}.

\section{Problem Formulation}

\noindent
Consider a network containing a set of vehicles $\bm{S}$ as service providers and a set of UAVs $\bm{U}$ as service requestors, the relevant constraints are listed below.

\setcounter{equation}{0}
\renewcommand{\theequation}{C\arabic{equation}}

\noindent
$\bullet$ \textbf{Available resource limitation} imposes restrictions on idle VMs for each SP: 
\begin{align}
\sum^{|\bm{U}|}_{m=1}{\sum^{|{\bm{V}}^{{\bm{u}}_{\bm{m}}}|}_{n=1}{x^{n, m}_k\le |\bm{{\mathcal{V}}_{k}}|}}, \forall s_k\in \bm{{\mathcal{R}}_{m}}.
\end{align}

\noindent
$\bullet$ \textbf{Soft opportunistic V2V connection} poses a soft constraint that if two connected components $v_{n, m}$ and $v_{n', m}$ of $u_m$ with weight $w^{u_m}_{n, n'}$ are mapped to different SPs $s_k$ and $s_{k'}$, the probability of the contact duration between $s_k$ and $s_{k'}$ being larger than $ \left|\frac{D_{n, m}}{r_{k, m}}-\frac{D_{n', m}}{r_{k', m}}\right|+w^{u_m}_{n, n'}$ should be greater than a threshold ${\alpha}_1 $ (${0<\alpha}_1<1$). Notably, $ \left|\frac{D_{n, m}}{r_{k, m}}-\frac{D_{n', m}}{r_{k', m}}\right|$ denotes the absolute value of the transmission time difference between component $v_{n, m}$ and $v_{n', m}$ and the order of components transmission is ignored.
\begin{align}
& e^{-\left( \left|\frac{D_{n, m}}{r_{k, m}}-\frac{D_{n', m}}{r_{k', m}}\right|+w^{u_m}_{n, n'}\right)\times w^s_{k, k'}}\ge {\alpha}_1,\notag\\
& \text{if } s_k\neq s_{k'} \text{ and } x^{n, m}_k\times x^{n', m}_{k'}=1, \notag \\
&\quad \forall e^{u_m}_{n, n'}\in {\bm{E}}^{{\bm{u}}_{\bm{m}}} \text{ and } \forall u_m\in \bm{U} \text{ and } \forall s_k, s_{k'}\in {\bm{\mathcal{R}_m}}.
\end{align} 

\noindent
$\bullet$ \textbf{Transmission power limitation} prevents the case where the total allocated power may exceed $u_m$'s upper limit $Q_m$.
\begin{align}
\sum^{|\bm{S}|}_{k=1}{\sum^{|{\bm{V}}^{{\bm{u}}_{\bm{m}}}|}_{n=1}{x^{n, m}_k\times q_{k, m}}}\le Q_m, \forall u_m\in \bm{U}
\end{align}

\noindent
$\bullet$ \textbf{UAV's coverage limitation} only allows each UAV to offload components to SPs within its communication radius. 
\begin{align}
q_{k, m}\triangleq 0, \forall u_m\in \bm{U}~\text{and}~s_k\notin \bm{{\mathcal{R}}_{m}}
\end{align}

\setcounter{equation}{8}
\renewcommand{\theequation}{\arabic{equation}}

For notational simplicity, let $\bm{x}={[x^{n, m}_k]}_{1\le k\le |\bm{S}|,1\le m\le |\bm{U}|,1\le n\le |{\bm{V}}^{{\bm{u}}_{\bm{m}}}|} $ denote the matrix of binary variable $x^{n, m}_k$, with size $|{\bm{V}}^{\bm{U}}|\times |\bm{S}|$, where $|{\bm{V}}^{\bm{U}}|$ and $|\bm{S}|$ indicate the total number of components and SPs, respectively. Let $\bm{q}={[q_{k, m}]}_{1\le k\le |\bm{S}|,1\le m\le |\bm{U}|} $ be the power allocation matrix of size $|\bm{U}|\times |\bm{S}|$, where $|\bm{U}|$ represents the number of UAVs in the system. Correspondingly, aiming to minimize the value of the objective function $\mathcal{F} (\bm{x}, \bm{q})$ given in~\eqref{eq9}, we formulate the proposed optimization problem of energy-aware graph task scheduling as $\bm{\mathcal{P}}$ in~\eqref{eq10}.
\begin{align}
\label{eq9} 
&\mathcal{F}(\bm{x}, \bm{q})={{\omega}_1\sum^{|\bm{U}|}_{m=1}{t_m} + {\omega}_2} \sum^{|\bm{U}|}_{m=1}{c_m}\notag\\
&+\frac{{\omega}_3}{2}\sum^{|\bm{S}|}_{k=1}\sum^{|\bm{S}|}_{k'=1}{\sum^{|\bm{U}|}_{m=1}{\sum^{|{\bm{V}}^{{\bm{u}}_{\bm{m}}}|}_{n=1}\sum^{|\bm{V^{u_m}}|}_{n'=1}{{\mathcal{C}}^{n, n', m}_{k, k'}}}},
\end{align} 

$\mathcal{F} (\bm{x}, \bm{q})$ stands for the weighted sum of task completion time, energy consumption and data exchange cost, where ${\omega}_1$, ${\omega}_2$ and ${\omega}_3$, represent the non-negative coefficients. $\sum^{|\bm{U}|}_{m=1}{t_m}$ and $\sum^{|\bm{U}|}_{m=1}{c_m}$ denote the overall task completion time and energy consumption of UAVs, respectively. $\frac{1}{2}\sum^{|\bm{S}|}_{k=1}{\sum^{|\bm{S}|}_{k'=1}}{\sum^{|\bm{U}|}_{m=1}{\sum^{|{\bm{V}}^{{\bm{u}}_{\bm{m}}}|}_{n=1}\sum^{|\bm{V^{u_m}}|}_{n'=1}{{\mathcal{C}}^{n, n', m}_{k, k'}}}}$ indicates the total data exchange cost among SPs in the VC, where the normalization factor ${1}/{2}$ is considered since the cost will be calculated twice due to the un-directed graph model. 
\begin{align} \label{eq10} 
\bm{\mathcal{P}}:~&{\mathop{\arg\min}_{\bm{x}, \bm{q}}\mathcal{F} (\bm{x}, \bm{q})} 
\end{align}
s.t.~~~~~~~~~~~~~~~~~~~~~~~~~~~(C1), (C2), (C3), (C4).

\vspace{0.15in}
Notably, $\bm{\mathcal{P}}$ stands for a non-trivial MINLP problem which is NP-hard with coupled binary variables $x^{n, m}_k\in \bm{x}$ and consecutive variables $q_{k, m}\in \bm{q}$, that both needed to be optimized. Moreover, constraint (C2) requires solving the subgraph isomorphism problem, which further poses challenges to the algorithm design~\cite{9,21,28,29}. In principle, solutions can be obtained through exhaustive search, which, however, is practically infeasible due to high complexity. For example, determining templates of mapping components of UAVs to SPs through exhaustive search results in high computational complexity of $O(2^{( |{\bm{V}}^{\bm{U}}|)\times \left(\sum^{|\bm{S}|}_{k=1}{|\bm{{\mathcal{V}}_{k}}|}\right)})$; and for each feasible mapping, the optimization problem of power allocation needs to be solved. Consequently, the system can rarely identify the optimal solutions to reconfigure the IoV extemporaneously, as the running time required to solve large and real-life network cases increases sharply with increasing vehicular and UAV's density. As a result, we propose an efficient decoupled approach for solving $\bm{\mathcal{P}}$ in the next section, which can offer a low computation complexity.

\section{Solving the Energy-aware Graph Task Allocation Problem: A Decoupled Approach}

\begin{table*}[b!]
{\small
\centering
\caption{Examples of the component exploration sequence, the related predecessors, and how a template is generated}
\setlength{\tabcolsep}{0.01mm}{
\begin{tabular}{*{7}{l}}
\hline\\[-2.9mm]\hline
\centering
Step &~~~~~~~~~~~~~~$\bm{N}$ &~~~~Predecessor & Component &~~~Candidate & Mapping &~~~~~~~~~~~~~~~~~Update \\
\hline
1 & $[]$ & -- & $F$ & $\{s_3,s_5,s_6\}$ & $F\to s_6$ & ${\mathcal{D}}^s(s_3)={\mathcal{D}}^s(s_5)=6$, ${\mathcal{D}}^s(s_6)=7$ \\
2 & $[F]$ & $\bm{Pred}(\bm{F}) = \{\}$ & $E$ & $\{s_3,s_5,s_6,s_7\}$ & $E\to s_7$ & ${\mathcal{D}}^s(s_6)=6$, ${\mathcal{D}}^s(s_7)=0$ \\
3 & $[F, E]$ & $\bm{Pred}(\bm{E}) = \{F\}$ & $H$ & $\{s_6\}$ & $H\to s_6$ & ${\mathcal{D}}^s(s_6)=0$, ${\mathcal{D}}^s(s_3)={\mathcal{D}}^s(s_5)=5$ \\
4 & $[F, E, H]$ & $\bm{Pred}(\bm{H}) = \{E, F\}$ & $G$ & $\{s_3,s_5\}$ & $G\to s_5$ & ${\mathcal{D}}^s(s_3)={\mathcal{D}}^s(s_5)=4$ \\
5 & $[F, E, H, G]$ & $\bm{Pred}(\bm{G}) = \{F\}$ & $I$ & $\{s_3,s_5\}$ & $I\to s_3$ & ${\mathcal{D}}^s(s_3)={\mathcal{D}}^s(s_5)=3$ \\
6 & $[F, E, H, G, I]$ & $\bm{Pred}(\bm{I}) = \{F, G\}$ & $A$ & $\{s_2,s_3,s_4,s_5\}$ & $A\to s_5$ & ${\mathcal{D}}^s(s_2)=6$, $ {\mathcal{D}}^s(s_3)= {\mathcal{D}}^s(s_4)=3$, ${\mathcal{D}}^s(s_5)=4$ \\
7 & $[F, E, H, G, I, A]$ & $\bm{Pred}(\bm{A}) = \{\}$ & $B$ & $\{s_2,s_3,s_4,s_5\}$ & $B\to s_2$ & ${\mathcal{D}}^s(s_2)=0$, $ {\mathcal{D}}^s(s_3)={\mathcal{D}}^s(s_4)=2$, ${\mathcal{D}}^s(s_5)=3$ \\
8 & $[F, E, H, G, I, A, B]$ & $\bm{Pred}(\bm{B}) = \{A\}$ & $C$ & $\{s_3,s_4,s_5\}$ & $C\to s_4$ & ${\mathcal{D}}^s(s_3)={\mathcal{D}}^s(s_5)=2$, $ {\mathcal{D}}^s(s_4)=0$ \\
9 & $[F, E, H, G, I, A, B, C]$ & $\bm{Pred}(\bm{C}) = \{A\}$ & $D$ & $\{s_3,s_5\}$ & $D\to s_5$ & ${\mathcal{D}}^s(s_3)=1$, $ {\mathcal{D}}^s(s_5)=0$ \\
10 & $[F, E, H, G, I, A, B, C, D]$ & $\bm{Pred}(\bm{D}) = \{A\}$ & \multicolumn{4}{l}{The related template $\bm{\mathcal{X}}: \{A, B, C, D, E, F, G, H, I\}\to \{s_5,s_2,s_4,s_5,s_7,s_6,s_5,s_6,s_3\}$} \\
\hline\\[-2.9mm]\hline
\end{tabular}}
\label{tab2}
}
\end{table*}

\noindent
The significance of preserving the structures of both the VC and the graph tasks complicates the simultaneous allocation of task components and transmission power among SPs. In this section, we propose an efficient approach by decoupling the template search problem from the power allocation problem, which mainly contains two stages. For the former stage, an efficient template search algorithm is proposed aiming to search for all the feasible mappings between the graph tasks and the SPs. Then, given the templates obtained from stage 1, a power allocation algorithm is presented via applying convex optimization techniques. 

\subsection{Stage 1: The Proposed Template Search Algorithm}

\noindent
The template stands for one of the key concerns in this paper, which is formalized as the search for all the subgraph isomorphisms~\cite{21} between the graph tasks and the VC. For analytical simplicity, let $\bm{X} = \{\bm{{\mathcal{X}}_{z}} |\bm{z}\in\{1,2,\cdots |\bm{X}|\}\}$ be the set of templates, where $\bm{{\mathcal{X}}_{z}}$ is the $z^{\rm{th}}$ template in set $\bm{X}$. Note that not every SP can be selected, we define $\widetilde{{\bm{S}}}_{\bm{z}}\subseteq \bm{S}$ as a subset of $\bm{S}$ associated with template $\bm{\mathcal{X}}_{\bm{z}}$. Specifically, let ${\tilde{s}}^z_k\in \widetilde{{\bm{S}}}_{\bm{z}}$ be the $k^{\text{th}}$ SP in set $\widetilde{{\bm{S}}}_{\bm{z}}$, and $\widetilde{\bm{\mathcal{V}}}^{\bm{z}}_{\bm{k}}$ be the related VM set of ${\tilde{s}}^z_k$. The edge and weight set corresponding to $\widetilde{{\bm{S}}}_{\bm{z}}$ are denoted as $\widetilde{{\bm{E}}}^{\bm{s}, \bm{z}} = \{{\tilde{e}}^{s, z}_{k, k'}| {\tilde{s}}^z_k, {\tilde{s}}^z_{k'}\in \widetilde{{\bm{S}}}_{\bm{z}}\}$ and $\widetilde{{\bm{W}}}^{\bm{s}, \bm{z}} = \{{\tilde{w}}^{s, z}_{k, k'}| {\tilde{s}}^z_k, {\tilde{s}}^z_{k'}\in \widetilde{{\bm{S}}}_{\bm{z}}\}$, where $\widetilde{{\bm{E}}}^{\bm{s}, \bm{z}}$ $\subseteq {\bm{E}}^{\bm{s}}$ and $\widetilde{{\bm{W}}}^{\bm{s}, \bm{z}}\subseteq {\bm{W}}^{\bm{s}}$. Thus, each template $\bm{{\mathcal{X}}_{z}}\bm{\in} \bm{X}$ can be represented as $\bm{{\mathcal{X}}_{z}} = \{x^{n, m}_k(z)|u_m\in \bm{U}, v_{n, m}{\bm{\in} \bm{V}}^{{\bm{u_m}}}, {\tilde{s}}^z_k\in \widetilde{{\bm{S}}}_{\bm{z}}\}$, where $x^{n, m}_k(z)$ indicates the assignment from component $v_{n, m}$ to SP ${\tilde{s}}^z_k\in \widetilde{{\bm{S}}}_{\bm{z}}$, in template $\bm{{\mathcal{X}}_{z}}$. The problem of searching for the templates is formulated as $\bm{\mathcal{P}_{1}}$ in~\eqref{eq11}. 
\begin{align} 
\label{eq11} 
\bm{{\mathcal{P}}_{1}}: \bm{X} 
\end{align} 
s.t.
\begin{align}
& \sum^{|\bm{U}|}_{m=1}{\sum^{|{\bm{V}}^{{\bm{u}}_{\bm{m}}}|}_{n=1}{x^{n, m}_k(z)\le}} |\widetilde{\bm{\mathcal{V}}}^{\bm{z}}_{\bm{k}}|, \forall {\tilde{s}}^z_k\in \widetilde{{\bm{S}}}_{\bm{z}} \text{ and } \forall \bm{{\mathcal{X}}_{z}}\in \bm{X}, \tag{C5}\\
&~~~~~~~~~~~~~~ x^{n, m}_k(z)\triangleq 0, \forall {\tilde{s}}^z_k\notin \bm{{\mathcal{R}}_{m}}, \bm{{\mathcal{X}}_{z}}\in \bm{X} \tag{C6}\\
& ~~~~~~~~~~~~~~ e^{-w^{u_m}_{n, n'}\times {\tilde{w}}^{s, z}_{k, k'}}\ge {\alpha}_2, \notag\\
& \text{ if } \exists e^{u_m}_{n, n'}\in {\bm{E}}^{{\bm{u}}_{\bm{m}}} \text{ and } x^{n, m}_k(z)\times x^{n', m}_{k'}(z)=1,\forall \bm{{\mathcal{X}}_{z}}\in \bm{X}. \tag{C7}
\end{align} 
Similar with (C1), constraint (C5) imposes restrictions on available VMs of each ${\tilde{s}}^z_k\in \widetilde{{\bm{S}}}_{\bm{z}}$. Constraint (C6) refers to the coverage limitation of each UAV, that is similar with (C4). Constraint (C7) poses a hard restriction that every template preserves the graph task structures; and a soft restriction which ensures that the probability of the contact duration between different SPs ${\tilde{s}}^z_k$ and ${\tilde{s}}^z_{k'}$ (processing connected components $v_{n, m}$ and $v_{n', m}$, respectively) being larger than $w^{u_m}_{n, n'}$ should be greater than a threshold ${\alpha}_2$ $(0<{\alpha}_2<1)$. Notably, the air-to-ground data transmission time is not concerned in $\bm{\mathcal{P}_{1}}$ owing to the unknown power allocation solution (which is introduced in the following Stage 2). 

The coverage overlaps may bring competitions among UAVs of available VMs. Thus, an efficient template search algorithm is proposed to solve $\bm{\mathcal{P}_{1}}$, while achieving conflict avoidance of re-occupations of the same VM among different UAVs. Our proposed algorithm is divided into two steps: \textbf{\textit{preprocess for obtaining the component exploration sequence}}, and \textbf{\textit{search for the templates}}. The first step stands for a preprocessing of the graph tasks, aiming to obtain an exploration sequence to determine the next candidate component during mapping. The second step aims to obtain the set of templates $\bm{X}$ under an efficient manner.

\textbf{Step 1. Preprocess for obtaining the component exploration sequence:} to prioritize the components that are more rare and constrained in graph tasks~\cite{21}, step 1 defines the order relationship by generating the component exploration sequence $\bm{N}$. Specifically, an exploration sequence $\bm{N}$ denotes a permutation of the components in set ${\bm{V}}^{\bm{U}}$, and is applied in step 2 to determine the next candidate component. In this paper, the rareness of a component $v_{n, m}{\bm{\in} \bm{V}}^{{\bm{u}}_{\bm{m}}}$ relies on the \textbf{\textit{degree of component}} ${\mathcal{D}}^c (v_{n, m})$, shown below.

\noindent
\textbf{Definition 1 (the degree of component ${\mathcal{D}}^c$)}: the degree ${\mathcal{D}}^c (v_{n, m})$ of component $v_{n, m}$ is calculated as the number of edges related to $v_{n, m}$ in graph task ${\bm{G}}^{{\bm{u}}_{\bm{m}}}$. In this paper, ${\mathcal{D}}^c (v_{n, m})$ stands for the rareness of component $v_{n, m}$, where the larger value of ${\mathcal{D}}^c$ represents a rarer component. 
\hfill $\blacksquare$ 

To preserve the structure of each graph task, the order of components in $\bm{N}$ is determined by computing the connections of a component with the components those are already in $\bm{N}$. Correspondingly, we define the \textbf{\textit{component mapping degree}} ${\mathcal{D}}^{map} (v_{n, m})$ of a component $v_{n, m}$ as shown below:

\noindent
\textbf{Definition 2 (the component mapping degree ${\mathcal{D}}^{map}$)}: the component mapping degree ${\mathcal{D}}^{map} (v_{n, m})$ of $v_{n, m}$ equals to the number of edges between component $v_{n, m}$ and all the components that are already inside $\bm{N}$.
 \hfill $\blacksquare$ 

\noindent
Therefore, the procedure of generating $\bm{N}$ firstly depends on the component mapping degree ${\mathcal{D}}^{map}$ of each component; if two or more components have the same ${\mathcal{D}}^{map}$, they are sorted according to the value of ${\mathcal{D}}^c$; if both ${\mathcal{D}}^{map}$ and ${\mathcal{D}}^c$ are equal, the choice is done randomly. Specifically, the component with the largest ${\mathcal{D}}^c$ is selected as the first component in $\bm{N}$; if more than one component have the same ${\mathcal{D}}^c$, randomly choose one of them to be the first in $\bm{N}$. The second column of \textbf{Table~2} depicts an example of the component exploration sequence related to the graph tasks given in {Fig}.~2(a), and the pseudocode of the preprocessing step is detailed in \textbf{Algorithm 1}. 

\textbf{Step 2. Search for the templates:} the exploration sequence obtained by step 1 enables a criterion of selecting the next candidate component during the template search procedure. Given an exploration sequence $\bm{N}$, the template set $\bm{X}$ can be obtained by searching for all the subgraph isomorphisms between the graph tasks and the VC, through step 2. To describe the interdependencies among components, we first define the \textbf{\textit{predecessor}} $\bm{Pred}({\bm{v}}_{\bm{n}, \bm{m}})$ of a component $v_{n, m}$ as below, several examples of which are given in \textbf{Table~2} (the third column).


\begin{algorithm}
{\small
\caption{Preprocess for obtaining the component exploration sequence (Step 1)}

\SetKwInOut{Input}{Input}\SetKwInOut{Output}{Output}

\Input{${\bm{G}}^{{\bm{u}}_{\bm{m}}} = \{{\bm{V}}^{{\bm{u}}_{\bm{m}}}, {\bm{E}}^{{\bm{u}}_{\bm{m}}}, {\bm{W}}^{{\bm{u}}_{\bm{m}}}\}, \forall u_m\bm{\in} \bm{U}$}

\Output{$\bm{N}$}

Initialization: $\bm{N}\leftarrow \bm{[]}$, calculate ${\mathcal{D}}^c (v_{n, m})$ for all $v_{n, m}\in {\bm{V}}^{{\bm{u}}_{\bm{m}}}$, $u_m\bm{\in} \bm{U}$

$\bm{N}\leftarrow $ the component with largest value of ${\mathcal{D}}^c$, if two or more components have the same ${\mathcal{D}}^c$, randomly choose one, 

\For{$i=1$ to $|{\bm{V}}^{\bm{U}}|- 1$}{
\For{all $v_{n, m}\notin \bm{N}$}{
calculate ${\mathcal{D}}^{map}(v_{n, m})$,

$\bm{N}\leftarrow \bm{N}\cup $ the component with the largest ${\mathcal{D}}^{map}(v_{n, m})$; if two or more components have the same value of ${\mathcal{D}}^{map}$, choose the one with the largest ${\mathcal{D}}^c$; if two or more components have the same values of both ${\mathcal{D}}^{map}$ and ${\mathcal{D}}^c$, randomly choose one,

}

$i=i+1$, 
}

\textbf{end algorithm}
}
\end{algorithm}

\noindent
\textbf{Definition 3 (the predecessor of component)}: the predecessor $\bm{Pred}({\bm{v}}_{\bm{n}, \bm{m}})$ of component $v_{n, m}$ is defined as a set of components that have one-hop connection with $v_{n, m}$, and located before $v_{n, m}$ in $\bm{N}$. Notably, some of the components may have no predecessor, such as the first component in sequence $\bm{N}$. 
\hfill $\blacksquare$ 

To preserve task structures, while ensuring the efficiency of the proposed template search algorithm, a component $v_{n, m}$ can only be mapped to a SP that can meet the related degree requirements, and the structure constraints (C6) and (C7) with the components in set $\bm{Pred}({\bm{v}}_{\bm{n}, \bm{m}})$. Correspondingly, we define the \textbf{\textit{current available degree}} ${\mathcal{D}}^s (s_k)$ of a SP $s_k \in {\mathcal{R}}_{\bm{m}}$, and the \textbf{\textit{candidate of a component}} $v_{n, m}$ as follows:

\noindent
\textbf{Definition 4 (the current available degree} ${\mathcal{D}}^s$ \textbf{of the SP)}: the current available degree ${\mathcal{D}}^s (s_k)$ of SP $s_k$ is calculated as the sum of the current available VMs of $s_k$, and that of the SPs which have one-hop connection with $s_k$. Notably, if there is no local VM available on $s_k$, ${\mathcal{D}}^s (s_k)\triangleq 0$. 
\hfill $\blacksquare$

\noindent
\textbf{Definition 5 (the candidate of component)}: a SP $s_k$ can be a candidate of component $v_{n, m}$ if and only if the following two conditions are both satisfied:

\noindent
Condition 1: ${\mathcal{D}}^s(s_k)\ge {\mathcal{D}}^c(v_{n, m})-{\mathcal{D}}^{map}(v_{n, m})$,

\noindent
Condition 2: map component $v_{n, m}$ to SP $s_k$ can meet all the edge and weight constraints with the components in set $\bm{Pred}({\bm{v}}_{\bm{n}, \bm{m}})$. 
\hfill ~~~~~~~~~~~~~~~~~~~~~~~~~$\blacksquare$ 


Take the graph tasks shown in {Fig}.~2(a) as example. Under the exploration sequence $\bm{N}=[F, E, H, G, I, A, B, C, D]$ shown in \textbf{Table~2}, the current available degree ${\mathcal{D}}^s (s_6)$ of $s_6$ before graph task allocation is ${\mathcal{D}}^s(s_6)={| \bm{\mathcal{V}_6}} |+{|\bm{\mathcal{V}_3}}|+{| \bm{\mathcal{V}_5}} |+{| \bm{\mathcal{V}_7}}|=2+2+3+1=8$. After allocate component $F$ to $s_6$, the current available degree of which is computed as ${\mathcal{D}}^s(s_6)=(2-1)+2+3+1=7$. Similarly, after component $E$ is assigned to $s_7$, the value of ${\mathcal{D}}^s(s_6)$ is updated as ${\mathcal{D}}^s(s_6)=1+2+3+(1-1)=6$ due to that the VM of $s_7$ has been occupied by $E$. Accordingly, ${\mathcal{D}}^s(s_7)=0$.

Given an exploration sequence $\bm{N}$, the major steps of the proposed template search algorithm are given in \textbf{Algorithm~2}. The main idea is to sequentially assign each component in set $\bm{N}$ to an unmapped candidate at a time, until all the templates are searched out. Notably, the computation complexity also relies on both the graph task and the VC structures. For example, consider the VC structure as a complete graph (there exist an edge between any two SPs), the computation complexity of the proposed algorithm may rise to the same level with exhaustive search. Thus, the proposed template search algorithm can provide a best computation complexity performance of $O( |{\bm{V}}^{\bm{U}}|)$, but a worst case equals to the exhaustive search algorithm. However, the proposed algorithm works on both the preprocess of graph tasks, and the selection of candidate SPs for each component, which greatly reduces the searching space during mapping. Owing to the flexible topologies of the graph tasks as well as the VCs in real-life applications and networks, we can make a weak assumption that in most cases, the proposed algorithm will offer a low computation complexity. 
 
\begin{algorithm}
{\small
\caption{Search for the templates (Step 2)}

\SetKwInOut{Input}{Input}\SetKwInOut{Output}{Output}

\Input{$\bm{N}, {\bm{G}}^{{\bm{u}}_{\bm{m}}} = \{{\bm{V}}^{{\bm{u}}_{\bm{m}}}, {\bm{E}}^{{\bm{u}}_{\bm{m}}}, {\bm{W}}^{{\bm{u}}_{\bm{m}}}\}, \forall u_m\bm{\in} \bm{U}$, $\bm{\mathcal{R}_{m}}, \forall u_m\bm{\in} \bm{U}, {\bm{G}}^{\bm{s}}$}

\Output{$\bm{X}$}

Initialization: get $\bm{Pred}({\bm{v}}_{\bm{n}, \bm{m}})$ for all $v_{n, m}\in {\bm{V}}^{\bm{U}}$, 

in each iteration $z$, 

\For{$i=1$ to $|\bm{N}|$}{

get the candidate set of $\bm{N}[i]$, $\%$$\bm{N}[i]$ denotes the $i^{\rm th}$ component in $\bm{N}$ 

assign $\bm{N}[i]$ to the first unmapped candidate SP, 

put the related component-SP pair into $\bm{{\mathcal{X}}_{z}}$, 

$i=i+1$, 
}

$\bm{X}\leftarrow {\bm{\mathcal{X}_z}}\cup \bm{X}$, 

until finish searching for all unmapped candidate SP of each component in each iteration, by following $\bm{N}$, 

\textbf{end algorithm when all templates are searched out}
}
\end{algorithm}

To better analyze our proposed template search algorithm, a walk through example is provided in \textbf{Table~2} (from the forth column to the seventh colunm), showing how a template is generated given the graph task and VC structures shown in Fig.~2(a), under the given exploration sequence $\bm{N}=[F, E, H, G, I, A, B, C, D]$.

\subsection{Stage 2: The Proposed Power Allocation Algorithm} 

\noindent
In this section, we study an effective transmission power allocation algorithm under the given templates obtained from stage 1. Owing to that the proposed algorithm works indistinguishably for various templates, and the transmission power allocation is independent among different UAVs, the indexes $z$ and $m$ referring to a unique template $\bm{{\mathcal{X}}_{z}}$ and UAV $u_m$ are  ignored. Hereafter, symbols $x^n_k$, $D_n$, $r_k$, $q_k$ and $Q$ are as the substitutions of $x^{n, m}_k (z)$, $D_{n, m}$, $r_{k, m}$, $q_{k, m}$ and $Q_m$; and ${\bm{V}}^{\bm{u}} = \{v_n| n\in \{1,2,\cdots, |{\bm{V}}^{\bm{u}}|\}\}$, ${\bm{E}}^{\bm{u}} = \{e^u_{n, n'}| n, n'\in \{1,2,\cdots, |{\bm{V}}^{\bm{u}}|\}, n\bm{\neq} n'\}$, and ${\bm{W}}^{\bm{u}} = \{w^u_{n, n'}| n, n'\in \{1,2,\cdots, |{\bm{V}}^{\bm{u}}|\}, n\bm{\neq} n'\}$ are utilized instead of ${\bm{V}}^{{\bm{u}}_{\bm{m}}}$, ${\bm{E}}^{{\bm{u}}_{\bm{m}}}$, and ${\bm{W}}^{{\bm{u}}_{\bm{m}}}$. Similarly, we apply $\widetilde{\bm{S}} = \{{\tilde{s}}_k| k\bm{\in} \{1,2,\cdots, |\widetilde{\bm{S}}|\}\}$, $\widetilde{{\bm{E}}}^{\bm{s}} = \{{\tilde{e}}^s_{k, k'}|{\tilde{s}}_k, {\tilde{s}}_{k'}\bm{\in} \widetilde{\bm{S}}, {\tilde{s}}_k\neq {\tilde{s}}_{k'}\}$ and $\widetilde{{\bm{W}}}^{\bm{s}} = \{{\tilde{w}}^s_{k, k'}|{\tilde{s}}_k, {\tilde{s}}_{k'}\bm{\in} \widetilde{\bm{S}}, {\tilde{s}}_k\neq {\tilde{s}}_{k'}\}$ to differentiate the SP set $\widetilde{{\bm{S}}}_{\bm{z}}$ of template $\bm{{\mathcal{X}}_{z}}$ from the total SP set $\bm{S}$, regardless of the index $z$ of a certain template. 

The power allocation in stage 2 is formulated as an optimization problem $\bm{\mathcal{P}_2}$ given by~\eqref{eq12}, where $\widetilde{\bm{q}} = {[q_k]}_{1\le k \le | \widetilde{\bm{S}}|} $ denotes the power allocation vector that needs to be optimized. Notably, the value of data exchange is fixed under any given template, and thus not considered in $\bm{\mathcal{P}_2}$.
\begin{align} \label{eq12}
\bm{\mathcal{P}_2}:& \mathop{\arg\min}\limits_{\widetilde{\bm{q}}}~{\omega}_1{\max {\left[\frac{D^{total}_k}{r_k}\right]}_{1\le k\le |\widetilde{\bm{S}}|}+{\omega}_2} \sum^{|\widetilde{\bm{S}}|}_{k=1}{\frac{D^{total}_k\times q_k}{r_k}}\notag \\
&\quad +{\omega}_1t^{exec}+{\omega}_2\ell
\end{align} 
s.t.
\begin{align}
& {\mathrm{e}}^{-\left( \left|\frac{D_n}{r_k}-\frac{D_{n'}}{r_{k'}}\right|\bm{+}w^u_{n, n'}\right)\times {\tilde{w}}^s_{k, k'}}\ge {\alpha}_1, \quad \forall {\tilde{s}}_k\neq {\tilde{s}}_{k'}~\text{and}\notag \\
& \quad x^n_k\times x^{n'}_{k'}=1 \text{ and } \forall e^u_{n, n'}\in {\bm{E}}^{\bm{u}} \text{ and }\forall{\tilde{s}}_k, {\tilde{s}}_{k'}\in \widetilde{\bm{S}}, \tag{C8}\\
& ~~~~~~~~~~~~~~~~~~~~~~~~~\sum^{|\widetilde{\bm{S}}|}_{k=1}{q_k}\le Q, \tag{C9}
\end{align}

\noindent
where $D^{total}_k\triangleq \sum^{|{\bm{V}}^{\bm{u}}|}_{m=1}{x^n_k\times D_n}$ denotes the total size of the data transmitted from the UAV to $\widetilde{s}_k$, which is known for any given template. Same with (C2), constraint (C8) provide a soft restriction on preserving the edges and weights of the graph task. Constraint (C9) guarantees that the upper limit of a UAV's transmission power will not be exceeded. Apparently, $\bm{\mathcal{P}_2}$ refers to a \textit{min-max} problem, which is challenging to be solved. For analytical simplicity, let $\bm{\mathcal{T}}={\left[\frac{D^{total}_k}{r_k}\right]}_{1\le k\le |\widetilde{\bm{S}}|}$ be a vector with length $ |\widetilde{\bm{S}}|$, and the first term ${\max {\left[\frac{D^{total}_k}{r_k}\right]}_{1\le k\le |\widetilde{\bm{S}}|}} $ of $\bm{\mathcal{P}_2}$ stands for vector's infinite-norm $\|\bm{{\mathcal{T}}}\|_{\infty}$. Thus, vector's $p$-norm $\|\bm{{\mathcal{T}}}\|_p ={\left(\sum^{|\widetilde{\bm{S}}|}_{k=1}{{\left(\frac{D^{total}_k}{r_k}\right)}^p}\right)}^{\frac{1}{p}}$ is applied to approximate the optimal solution of $\bm{\mathcal{P}_2}$.
\begin{align} 
\label{eq13} 
{\|\bm{{\mathcal{T}}}}\|_{p\uparrow} \rightarrow \|\bm{{\mathcal{T}}}\|_{\infty}
\end{align} 
Apparently, \eqref{eq13} indicates that a larger value of $p$ enables the value of 
$\|\bm{\mathcal{T}} \|_p$ to approach $\|\bm{{\mathcal{T}}}\|_{\infty} $, and correspondingly brings a lower peak value of vector $\bm{\mathcal{T}}$. To facilitate the analysis, we consider $\|\bm{{\mathcal{T}}}\|^p_p =\sum^{|\overline{\bm{S}}|}_{k=1}{{\left(\frac{D^{total}_k}{r_k}\right)}^p}$ ($p\ge 1$) as the substitution of ${\max {\left[\frac{D^{total}_k}{r_k}\right]}_{1\le k\le |\widetilde{\bm{S}}|}} $. Thus, $\bm{\mathcal{P}_2}$ is rewritten as $\bm{\mathcal{P}_3}$ shown in~\eqref{eq14}, where constants ${\omega}_1t^{exec}$ and ${\omega}_2\ell $ can be ignored. 
\begin{align} \label{eq14} 
\bm{{\mathcal{P}_3}}: \mathop{\arg\min}\limits _{\widetilde{\bm{q}}} {\omega}_1\sum^{|\widetilde{\bm{S}}|}_{k=1}{{\left(\frac{D^{total}_k}{r_k}\right)}^p} +{\omega}_2\sum^{|\widetilde{\bm{S}}|}_{k=1}{\frac{D^{total}_k\times q_k}{r_k}}
\end{align} 
s.t. \centerline{(C8), (C9).}

Although we may concentrate on obtaining the optimal solution of $\bm{\mathcal{P}_3}$, there still remaining difficulties featured by non-convexity shown in the following lemma.


\noindent
\textit{Lemma 1}: $\bm{\mathcal{P}_3}$ represents a non-convex optimization problem.

\noindent
\textit{Proof}. According to equation~\eqref{eq6}, the data transmission rate $r_k$ is a function of $q_k$. Thus, the proof can be obtained by verifying the concavity of $\frac{q_k}{r_k (q_k)}$~\cite{41}, which makes $\bm{\mathcal{P}_3}$ non-convex. 
\hfill $\blacksquare$ 

In consequence, the \textit{change-of-variable} technique is applied to transform $\bm{{\mathcal{P}_3}}$ into a convex optimization problem $\bm{\mathcal{P}}_{\bm{4}}$, by introducing a set of substitutive variables $\bm{\rho}=\{\rho_k|{\rho_k=1}/{r_k}\}$.
\begin{align} 
\label{eq15} 
\bm{{\mathcal{P}}_{4}}: \mathop{\arg\min}\limits_{\bm{\rho}} \sum^{|\widetilde{\bm{S}}|}_{k=1}{\left({\mathcal{W}}_{1,k}{\left(\rho_k\right)}^p+{\mathcal{W}}_{2,k}\rho_k\left(2^{\frac{1}{B\times \rho_k}}-1\right)\right)} 
\end{align}
s.t.
\begin{align}
&~~~~~~~~~~~~~~\sum^{|\widetilde{\bm{S}}|}_{k=1}{\frac{1}{g_k}\times 2^{\frac{1}{B\times \rho_k}}-C^*\le 0}, \tag{C10}\\
&|\rho_k-\rho_{k'}|+C^{n, n'}_{k, k'}\le 0, \forall {\tilde{s}}_k\neq {\tilde{s}}_{k'} \text{ and } x^n_k\times x^{n'}_{k'}=1 \text{ and } \notag \\
&\forall e^u_{n, n'}\in {\bm{E}}^{\bm{u}} \text{ and } \forall {\tilde{s}}_k, {\tilde{s}}_{k'}\in \widetilde{\bm{S}}, \tag{C11}
\end{align}

\noindent
where $ {\mathcal{W}}_{1,k}\triangleq {\omega}_1{(D^{total}_k)}^p$, ${\mathcal{W}}_{2,k}\triangleq \frac{{\omega}_2D^{total}_kN_0}{g_k}$, and $C^*\triangleq \frac{Q}{N_0}+\sum^{|\widetilde{\bm{S}}|}_{k=1}{\frac{1}{g_k}}$, which are constants. $C^{n, n'}_{k, k'}=\frac{{\ln {\alpha} _1} +w^u_{n, n'}\times {\tilde{w}}^s_{k, k'}}{D\times {\tilde{w}}^s_{k, k'}}$ denotes the parameter when two connected components $v_n$ and $v_{n'}$ are mapped to different SPs ${\tilde{s}}_k$ and ${\tilde{s}}_{k'}$. For any given template, the value of each $C^{n, n'}_{k, k'}$ is also fixed (namely, each $C^{n, n'}_{k, k'}$ is a constant). Thus, (C11) can be rewritten as (C12) owning to $C^{n, n'}_{k, k'}\le 0$.
\begin{align}
& {(\rho_k-\rho_{k'})}^2-{(C^{n, n'}_{k, k'})}^2\le 0,\forall {\tilde{s}}_k\neq {\tilde{s}}_{k'} \text{ and } x^n_k\times x^{n'}_{k'}=1 \text{ and } \notag \\
&\forall e^u_{n, n'}\in {\bm{E}}^{\bm{u}} \text{ and } \forall {\tilde{s}}_k, {\tilde{s}}_{k'}\in \widetilde{\bm{S}} \tag{C12}
\end{align}

Due to that there may exists multiple pairs of connected components being mapped to different SPs, constraint (C12) is thus represented as a inequality constraints set $\bm{C}$, shown as (C13):
\begin{align}
& \bm{C}=\bigg\{f_i\left(\rho_k, \rho_{k'}, C^{n, n'}_{k, k'}\right)\le 0 |i\in \{1,\cdots, |\bm{C}|\}, \forall {\tilde{s}}_k\neq {\tilde{s}}_{k'}\notag \\
& \quad \text{and } x^n_k\times x^{n'}_{k'}=1 \text{ and } \forall e^u_{n, n'}\in {\bm{E}}^{\bm{u}} \text{ and } \forall {\tilde{s}}_k, {\tilde{s}}_{k'}\in \widetilde{\bm{S}}\bigg\}, \tag{C13}
\end{align}
where $i$ denotes the index of the inequality, $f_i(\rho_k, \rho_{k'}, C^{n, n'}_{k, k'})$ $={(\rho_k-\rho_{k'})}^2-{(C^{n, n'}_{k, k'})}^2$ and each pair of $n, n', k, k'$ can be known under a given template. $|\bm{C}|$ denotes the number of inequalities in set $\bm{C}$, which equals to the number of connected component pairs been mapped to different SPs. Notably, various templates\footnote{For example, the constraint set $\bm{C}$ of UAV 1 shown in {Fig}.~1 (a) contains two inequations: $ f_{1}(q_{s_{2}}, q_{s_{5}}, C^\text{AB}_{s_{2}s_{5}})={(q_{s_{2}}-q_{s_{5}})}^{2}-{(C^\text{AB}_{s_{2}s_{5}})}^{2}\le 0$, and $ f_{2}(q_{s_{4}}, q_{s_{5}}, C^\text{AC}_{s_{4}s_{5}})={(q_{s_{4}}-q_{s_{5}})}^{2}-{(C^\text{AC}_{s_{4}s_{5}})}^{2}\le 0$. Here, $A$, $B$, $C$ and $D$ are used to indicate components of the graph task for notational simplicity.} can have different $\bm{C}$. Correspondingly, the problem $\bm{\mathcal{P}_4}$ represents a convex optimization problem, as proved in \textit{Lemma 2}.

\noindent
\textit{Lemma 2:} $\bm{\mathcal{P}}_{\bm{4}}$ represents a convex optimization problem.

\noindent
\textit{Proof.} Let function $y(\rho)=y_1(\rho)+y_2(\rho)$, where $y_1(\rho)={\mathcal{W}}_{1,k}{(\rho)}^p$ and $y_2(\rho)={\mathcal{W}}_{2,k}\rho(2^{\frac{1}{B\times \rho}}-1)$. The second-order derivative of $ y(\rho)$ can be given by:
\begin{align} 
\label{eq16} 
\frac{d^2y(\rho)}{d^2\rho}=y''_1(\rho)+y''_2(\rho), 
\end{align} 
where $y''_1(\rho)=p(p-1){\mathcal{W}}_1\rho^{p-2}\ge 0$, and $y'_2(\rho)$ is calculated as:
\begin{align} 
\label{eq17} 
y'_2(\rho)& = \frac{d^1y_2(\rho)}{d^1\rho}  \\
& ={({\mathcal{W}}_{2,k}\rho)}'\times (2^{\frac{1}{B\times \rho}}-1)+{\mathcal{W}}_{2,k}\rho\times {(2^{\frac{1}{B\times \rho}}-1)}',  \notag
\end{align} 
where ${({\mathcal{W}}_{2,k}\rho)}'={\mathcal{W}}_{2,k}$ and ${(2^{\frac{1}{B\times \rho}}-1)}'=-\frac{{\ln 2}}{B}{\times 2}^{\frac{1}{B\times \rho}}\times \rho^{-2}$. Thus, we have $y''_2(\rho)$ shown in~\eqref{eq18}.
\begin{align} 
\label{eq18} 
y''_2(\rho)=\frac{d^2y_2(\rho)}{d^2\rho}=2^{\frac{1}{B\times \rho}}\times \frac{{\ln}^22}{B^2{\times \rho}^2}
\end{align} 

\noindent
Apparently, $\frac{d^2y(\rho)}{d^2\rho}>0, \forall \rho>0$, and $y(\rho)$ represents a convex function of variable $\rho$. Since $\bm{\mathcal{P}_4}$ aims at minimizing a summation of convex functions $y(\rho_k)$, where the constraint (C10) is convex with ${(2^{\frac{1}{B\times \rho}})}''>0 (\forall \rho>0$). Moreover, each inequality constraint in (C13) is convex, $\bm{\mathcal{P}_{4}}$ is proved to be a convex optimization problem. \hfill $\blacksquare$ 

In consequence, the power allocation vector can be obtained numerically by using convex optimization solvers such as MATLAB function fmincon and CVX~\cite{42}.

\section{Performance Evaluation}

\noindent
This section presents numerical results that illustrate the validity of the proposed decoupled approach (abbreviate to ``Proposed'' for simplicity). In the following, the performance of the proposed template search and power allocation algorithms comparing with several baseline methods, are analyzed in detail. Moreover, various problem sizes are investigated considering different graph task types and VC structures, as well as various numbers of UAVs and SPs.

\subsection{Simulation Setup}

\noindent
We consider a simulation space of $1000\text{m}\times1000\text{m}\times100\text{m}$ (length$\times$width$\times$height), wherein the height of the UAV is randomly chosen from $80\text{m}$ to $100\text{m}$. The graph task types considered in this simulation are depicted in Fig. 2(b). The monotone increasing function $f(pl(d_{k, k'}))=0.15\times pl(d_{k, k'})+0.001$ is applied to determine the data exchange cost among different SPs. The main simulation parameters are randomly obtained from the following intervals: $D \in [500\text{Kb}, 600\text{Kb}]$, $Q\in[1.5\text{Watts}, 2\text{Watts}]$, $N_0 \in[4\text{mWatts}, 5\text{mWatts}]$, $B\in [10\text{MHz}, 12\text{MHz}]$, $\omega_1=\omega_2=\omega_3=1/3$, $w_{n,n'}^{u_m}\in{[0.1, 0.3]}$, $t^{exec}\in[0.1, 0.2]$, $w^s_{k, k'}\in [0.05,0.06]$ for small problem size cases, and $w^s_{k, k'}\in [0.01,0.02]$ for large problem size cases.

\begin{figure}[b!]
\centerline{\includegraphics[width=.9\linewidth]{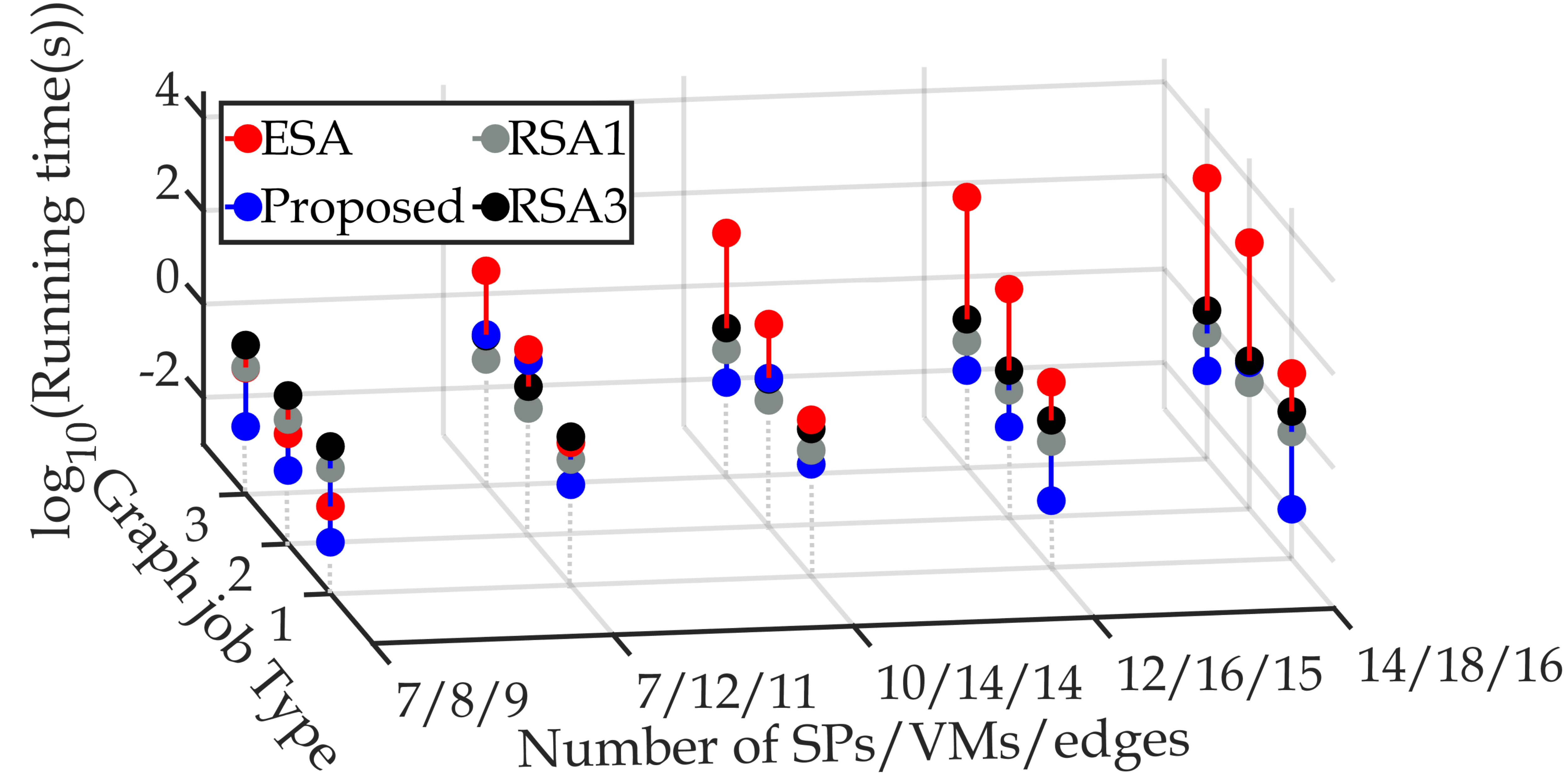}}
\caption{Performance comparisons of running time in small problem size cases.}
\end{figure}

\begin{figure*}[h!t]
\centering
\subfigure[]{\includegraphics[width=.325\linewidth]{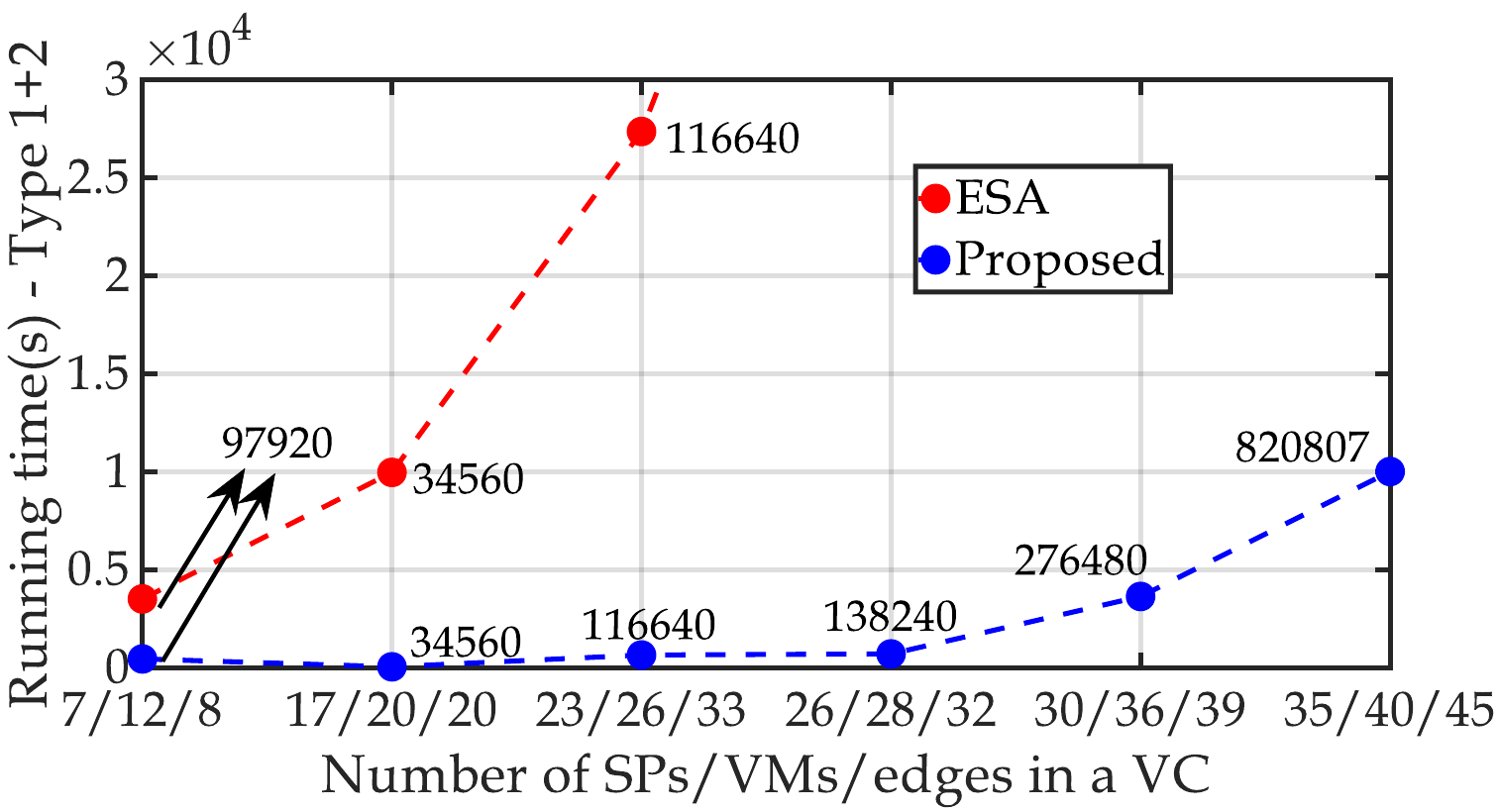}}
\subfigure[]{\includegraphics[width=.325\linewidth]{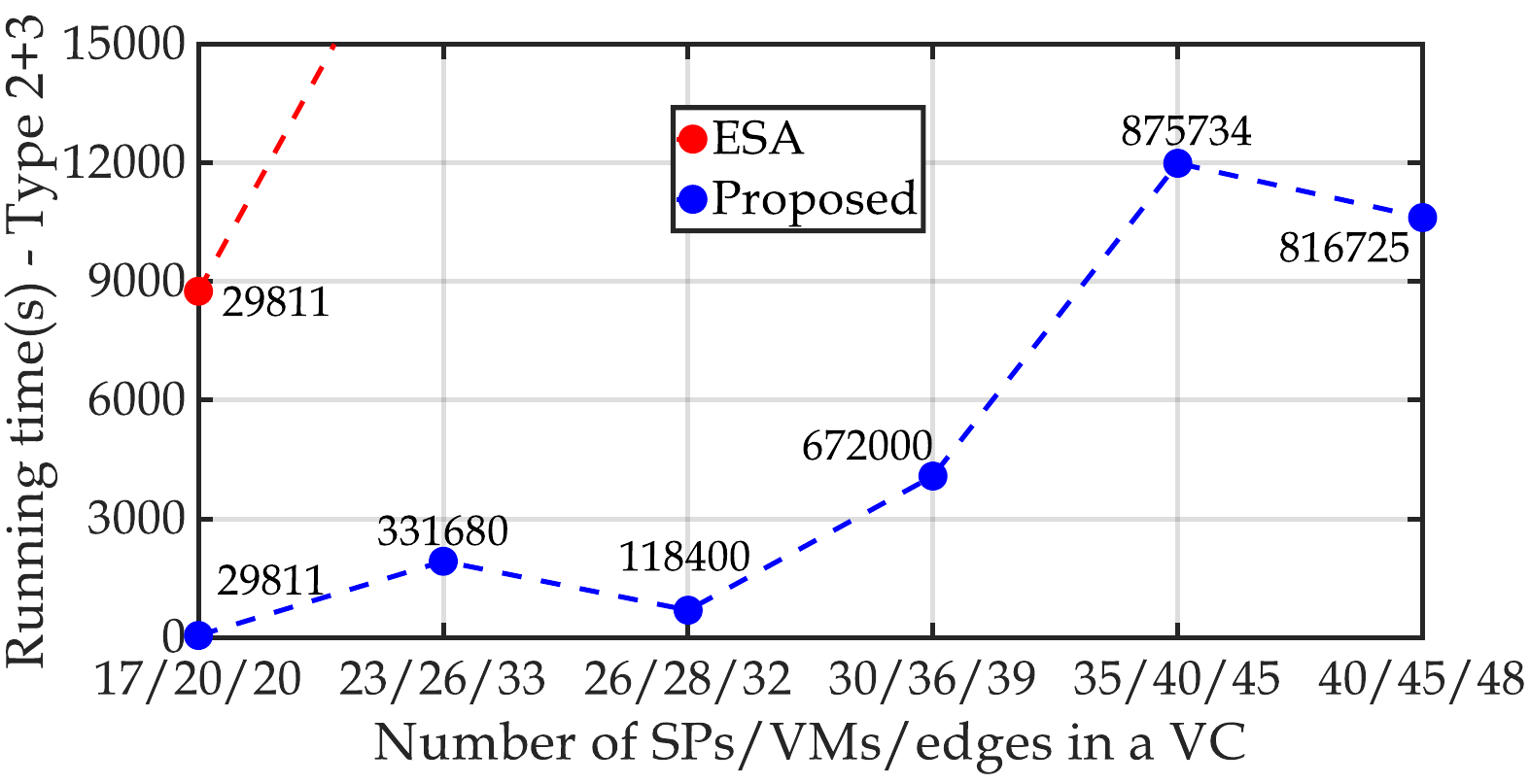}}
\subfigure[]{\includegraphics[width=.325\linewidth]{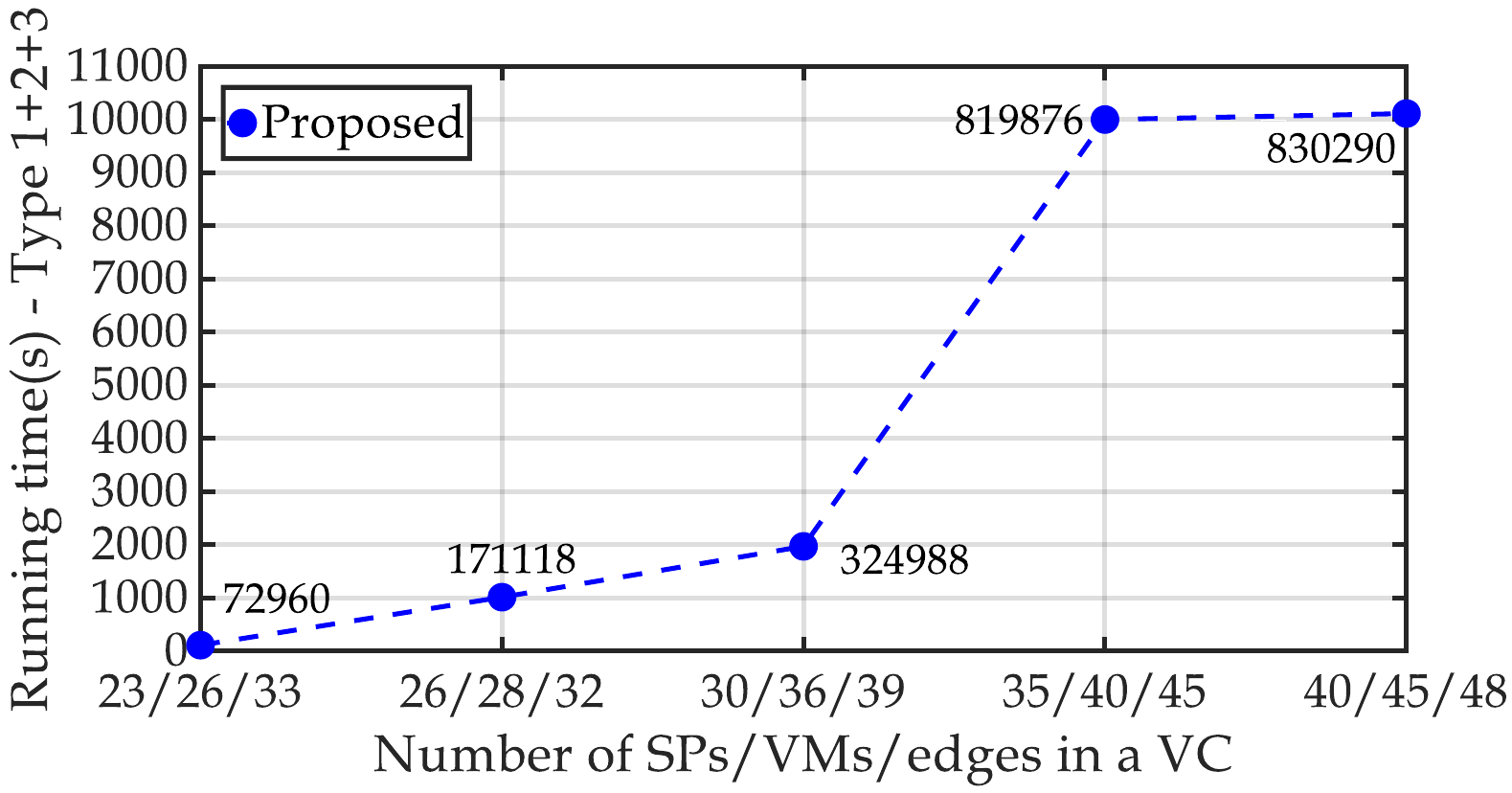}}
\caption{Performance comparisons of running time and templates count in large problem size cases: a). considering graph task type 1 and type 2; b). considering graph task type 2 and type 3; c). considering graph task type 1, type 2 and type 3.}
\end{figure*}
\subsection{Performance Comparisons of Template Searching}

\noindent
This section presents performance comparisons of the running time, and the number of templates (use ``templates count'' instead for notational simplicity) between the proposed template search algorithm and baseline methods listed below:

\noindent
1) \textbf{Exhaustive search algorithm (ESA)~\cite{8}}: check through all the mappings between the graph tasks and the VC structure, where the feasible ones are regarded as templates. 

\noindent
2) \textbf{Random search algorithm (RSA)}: randomly select a component and randomly match it to a SP at a time, until find a template. Notably, we consider different number of iterations: 10000 (RSA1), 20000 (RSA2), and 30000 (RSA3), to better demonstrate the performance evaluation.

\begin{figure*}[t!]
\centering
\subfigure[]{\includegraphics[width=.302\linewidth]{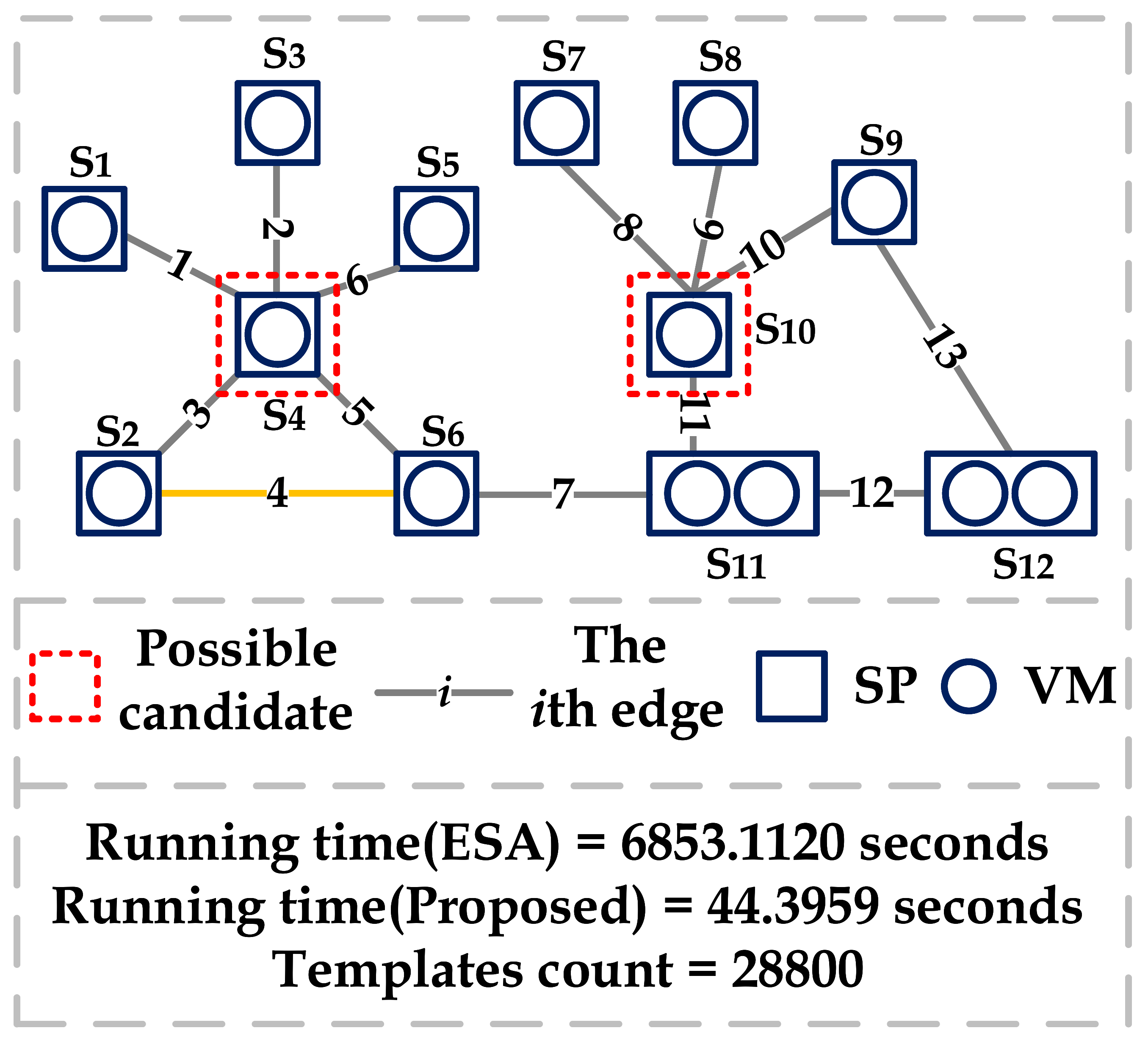}}
\subfigure[]{\includegraphics[width=.30\linewidth]{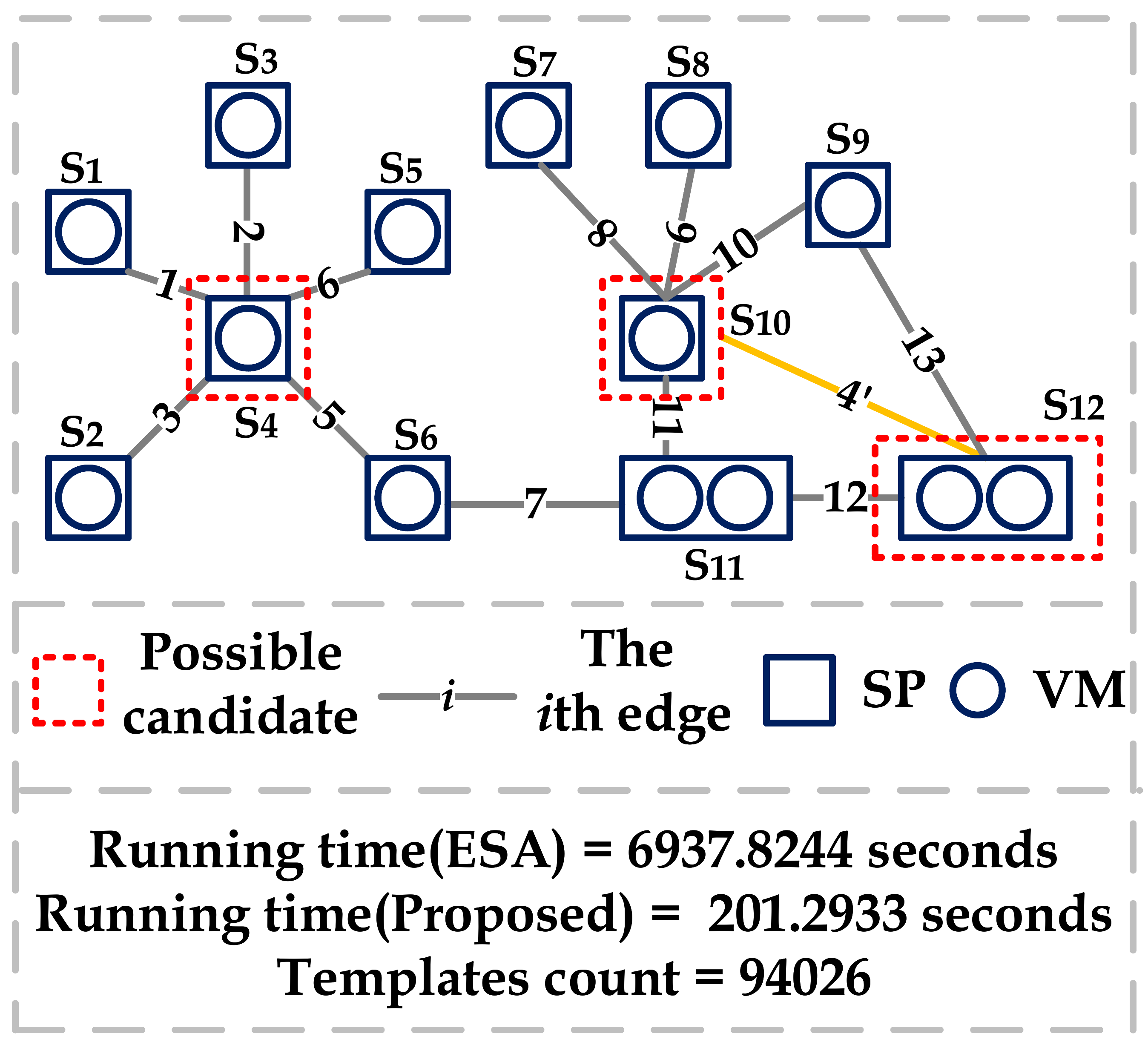}}
\subfigure[]{\includegraphics[width=.30\linewidth]{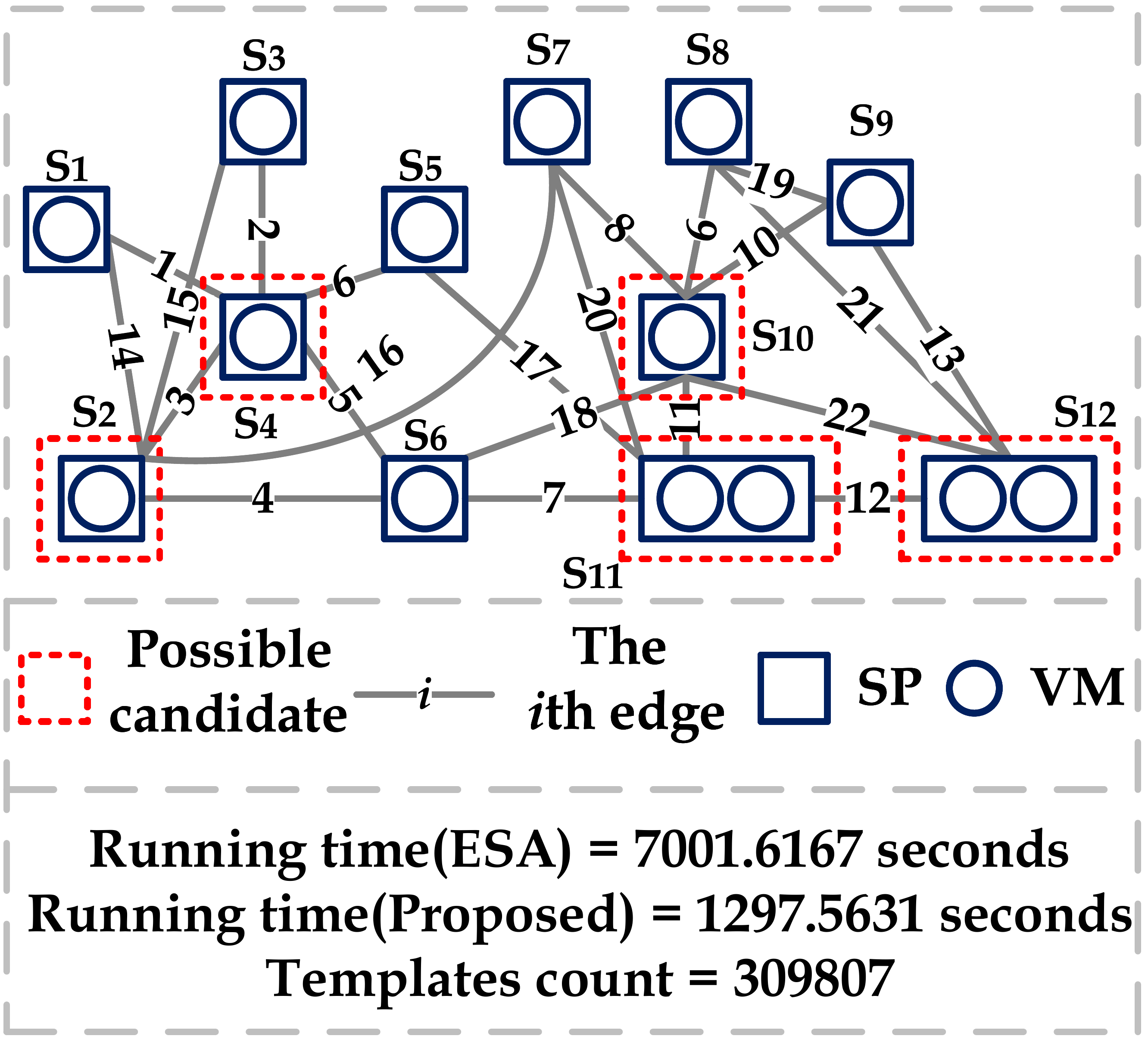}}
\caption{Running time performance comparisons between the VC structures with the same number of SPs and VMs (including two UAVs, each UAV carries a graph task type 2): a). a VC contains 12 SPs/14 VMs/13 edges, where the $4^{\rm{th}}$ edge locates between $s_2$ and $s_6$; b). a VC contains 12 SPs/14 VMs/13 edges, where the $4^{\rm{th}}$ edge locates between $s_{10}$ and $s_{12}$; c). a VC contains 12 SPs/14 VMs/22 edges.}
\end{figure*}

\begin{table}[h!]
{\footnotesize
\center
\caption{Performance comparisons of the templates count in small problem size~cases}
\tabcolsep 4pt
\setlength{\tabcolsep}{0.75mm}{
\begin{tabular}{p{1.2cm}<{\raggedright}*{6}{l}}
\hline\\[-2.9mm]\hline
\multicolumn{2}{c}{SPs/VMS/edges} & 7/8/9 & 7/12/11 & 10/14/14 & 12/16/15 & 14/18/16 \\
\hline
\multirow{5}{1.2cm}{Graph task type 1} & ESA & 16 & 1112 & 216 & 36 & 24 \\
& Proposed & 16 & 1112 & 216 & 36 & 24 \\
& RSA1 & 1 & 9 & 5 & 3 & 1 \\
& RSA2 & 2 & 13 & 9 & 5 & 1 \\
& RSA3 & 2 & 32 & 17 & 6 & 3 \\
\hline
\multirow{5}{1.2cm}{Graph task type 2} & ESA & 120 & 20400 & 7320 & 360 & 6840 \\
& Proposed & 120 & 20400 & 7320 & 360 & 6840 \\
& RSA1 & 5 & 37 & 11 & 9 & 14 \\
& RSA2 & 9 & 50 & 26 & 11 & 25 \\
& RSA3 & 16 & 59 & 30 & 17 & 37 \\
\hline
\multirow{5}{1.2cm}{Graph task type 3} & ESA & 4 & 7840 & 528 & 2344 & 284 \\
& Proposed & 4 & 7840 & 528 & 2344 & 284 \\
& RSA1 & 0 & 9 & 6 & 11 & 6 \\
& RSA2 & 0 & 19 & 10 & 13 & 8 \\
& RSA3 & 1 & 27 & 12 & 18 & 11 \\
\hline\\[-2.9mm]\hline
\end{tabular}
}
\label{tab3}
}
\end{table}

The running time and the template count performance comparisons in small problem size containing one UAV and a couple of SPs, are shown in {Fig}.~3 and {\textbf{Table~3}}, respectively. The 10-based logarithm representation is applied in {Fig}.~3, since the gap between the running time of various algorithms becomes large as the graph task and VC structures become more complicated (e.g., upon increasing the number of components, SPs/VMs and edges). As can be seen in {Fig}.~3, the running time of ESA rises sharply owing to that ESA relies heavily on the number of components, available VMs and the complexity of graph task as well as VC structures, which makes the ESA inappropriate for fast-changing and large-scale networks. Comparatively, the running time of the proposed algorithm remains approximately at a certain order of magnitude of 10$^{-2}$ $\sim$ 1~seconds when considering small problem sizes. Since the running time of the RSA mainly depends on the number of iterations, the performance of which remains slightly fluctuant near 0.5~seconds (RSA1), 1~seconds (RSA2) and 1.5~seconds (RSA3). \textbf{Table~3} presents the comparisons of templates count between different algorithms in small problem size cases. Apparently, our proposed algorithm can search for all the subgraph isomorphisms between the graph tasks and the VC, while offering a much lower computation complexity than ESA according to {Fig}.~3. The templates count of RSA under different numbers of iteration are far less than that of the proposed algorithm due to the randomness. In fact, failures and repetitions of mappings are common during the RSA procedure, owing to the structure preservation constraint. 

\begin{figure*}[t!]
\centering
\subfigure[]{\includegraphics[width=.315\linewidth]{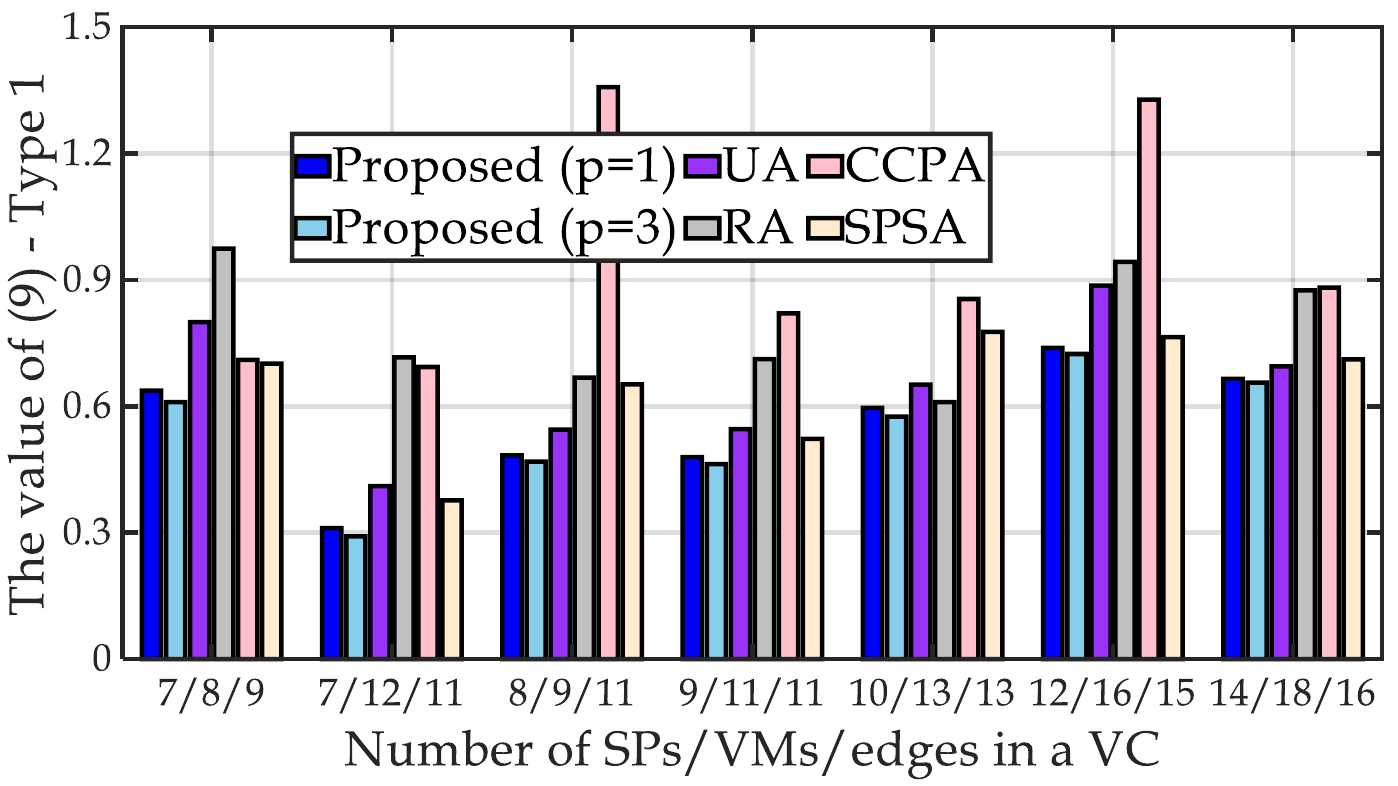}}
\subfigure[]{\includegraphics[width=.315\linewidth]{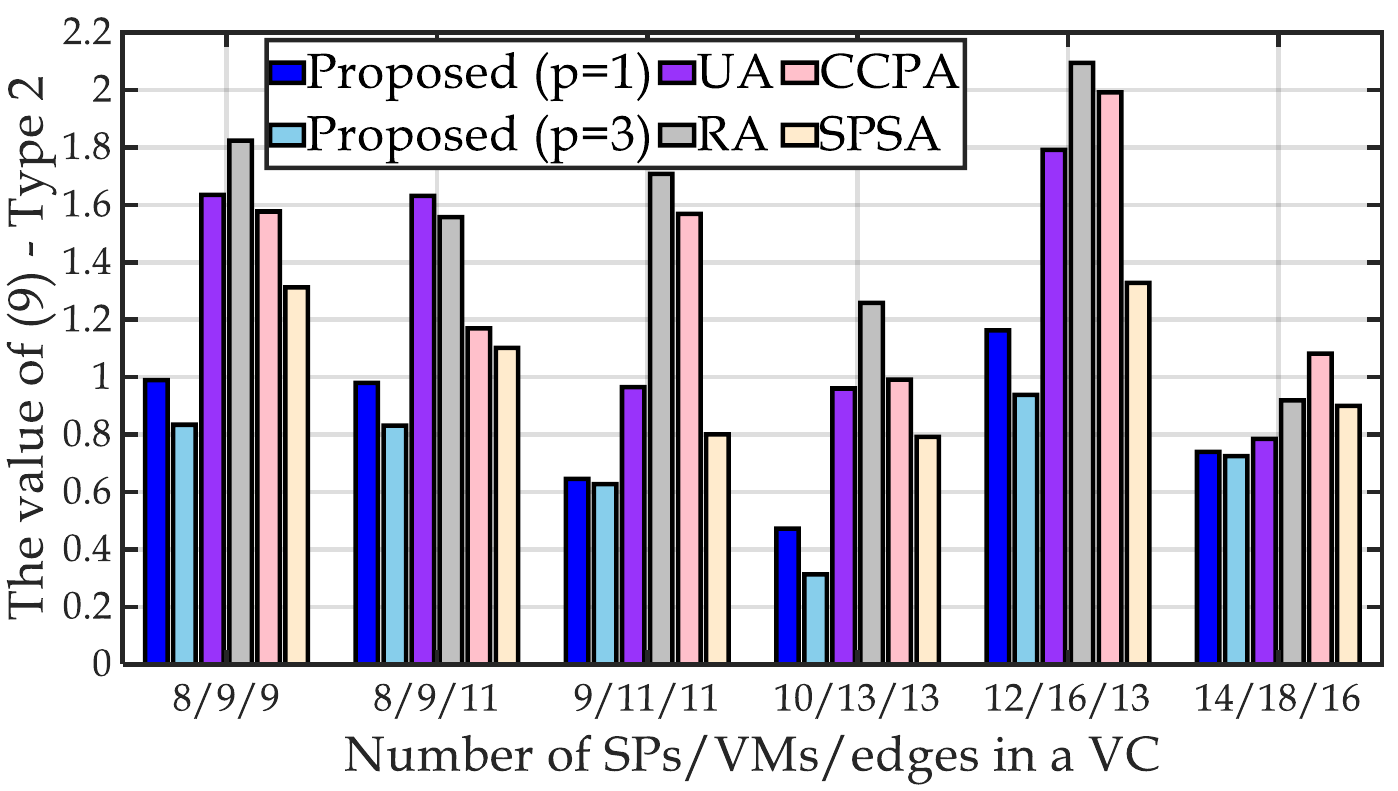}}
\subfigure[]{\includegraphics[width=.315\linewidth]{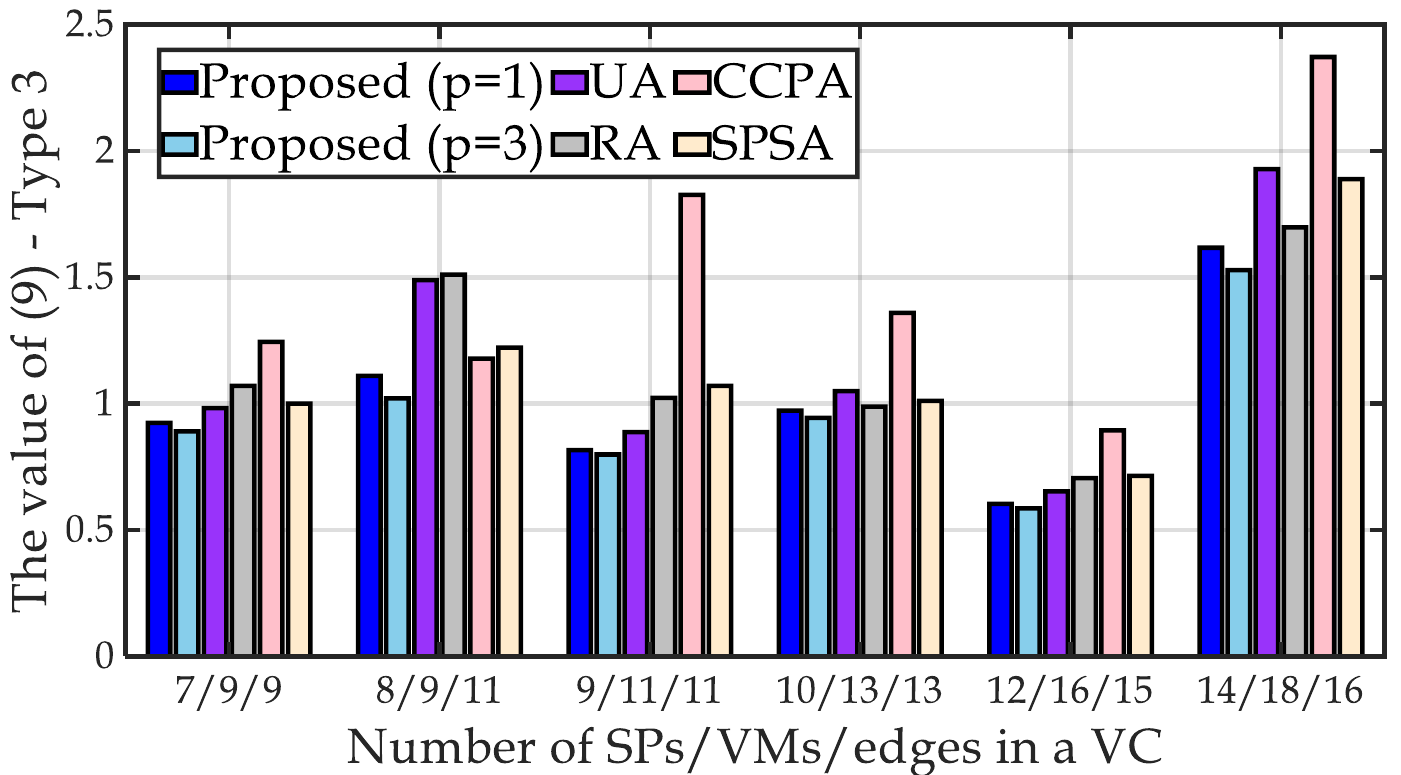}}
\subfigure[]{\includegraphics[width=.315\linewidth]{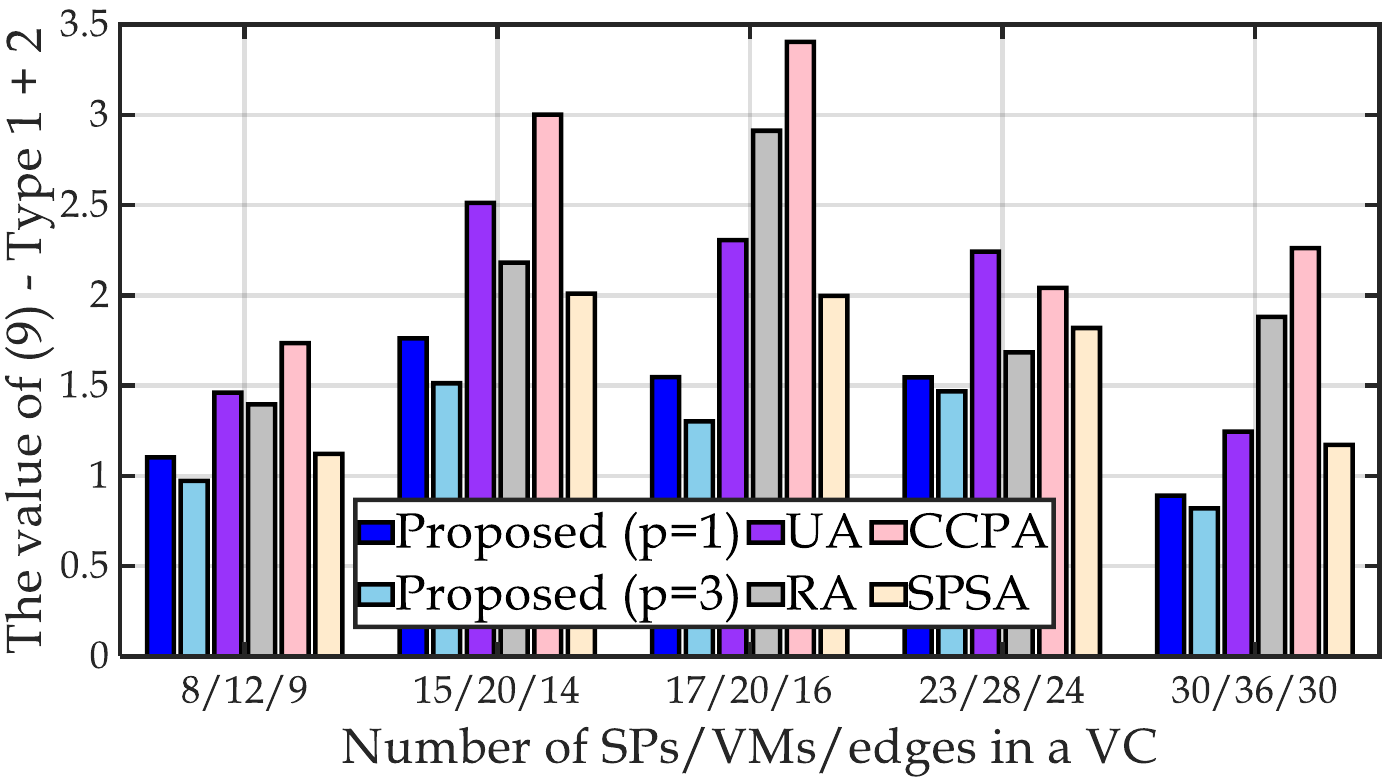}}
\subfigure[]{\includegraphics[width=.315\linewidth]{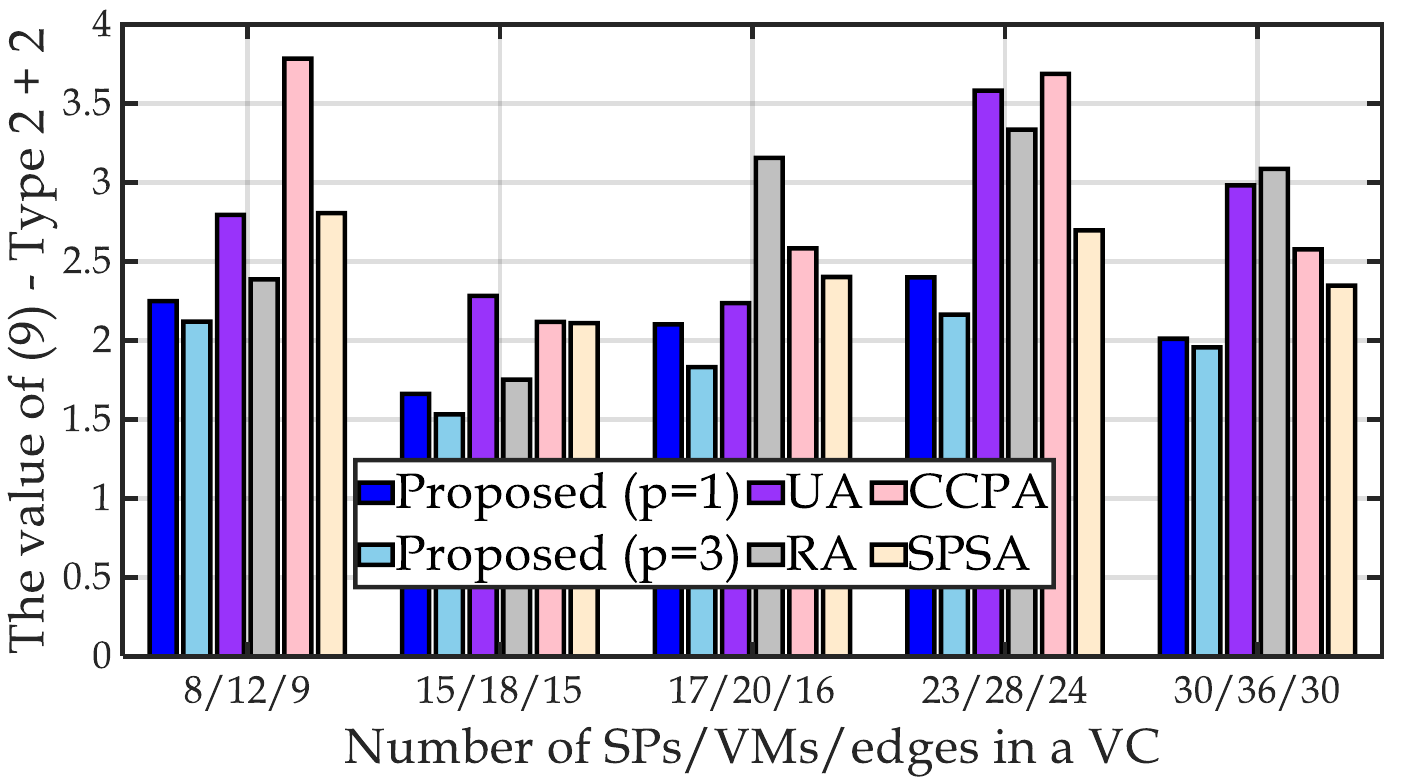}}
\subfigure[]{\includegraphics[width=.315\linewidth]{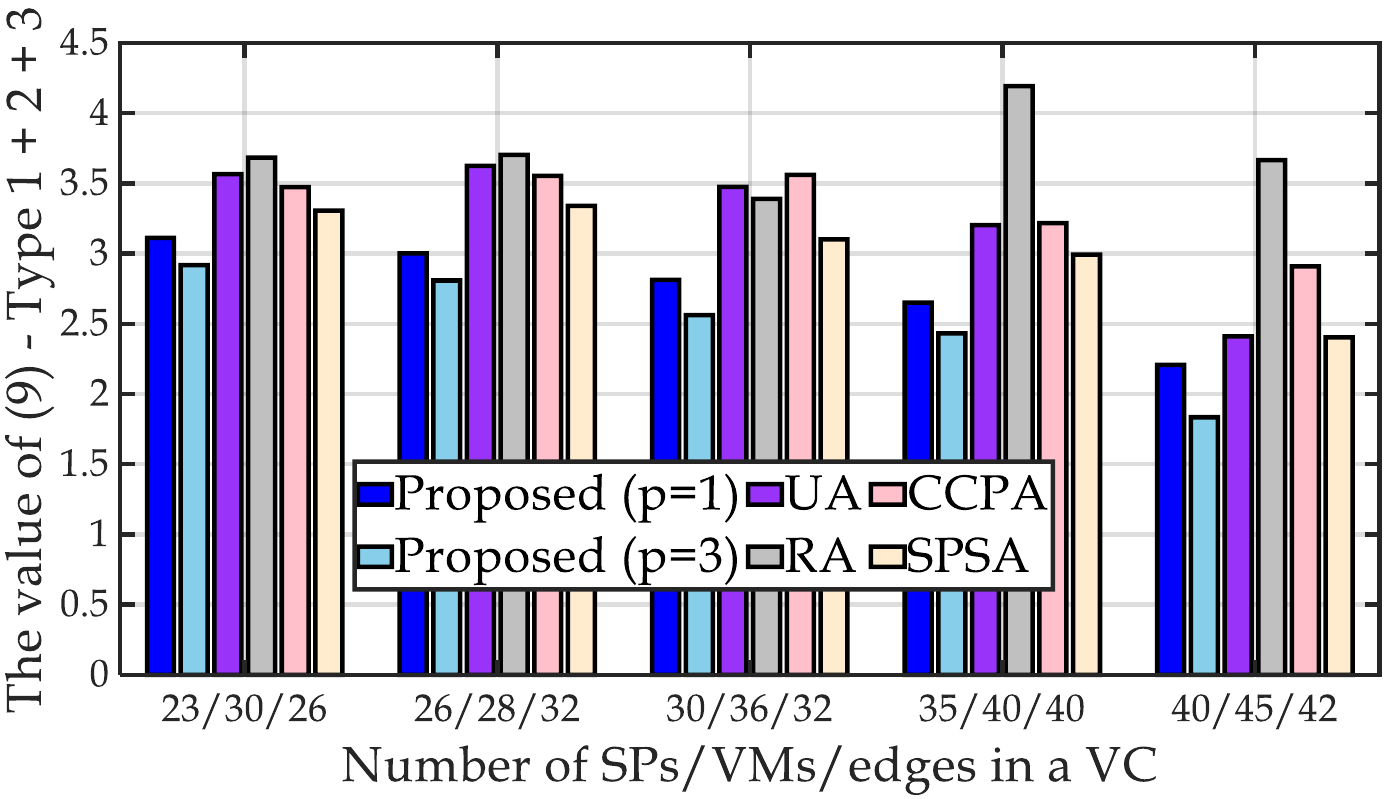}}
\caption{Performance comparison of the value of $\mathcal{F}(\bm{x},\bm{q})$ given in (9) considering various problem sizes.}
\end{figure*}
Fig.~4 presents the performance comparisons of running time and templates count in large problem size cases with multiple UAVs, increasing number of SPs and VMs, as well as more complicated VC structures (e.g., more edges between SPs). Here, we do not consider the baseline method RSA, since the running time of which relies mainly on the number of iterations, and the corresponding templates count is heavily influenced by the randomness factor. Notably, the number besides each mark in {Fig}.~4 indicates the number of templates related to the VC. {Fig}.~4(a) and {Fig}.~4(b) show the performance comparisons between ESA and the proposed template algorithm considering various scenarios with two graph tasks, and increasing number of SPs, VMs, and edges. As can be seen from these two figures, the proposed algorithm can reach the same templates count with ESA, while offering a much lower running time. Due to the too large running time of ESA, we only present the performance of the proposed algorithm in {Fig}.~4(c), under cases considering three graph tasks and more SPs/VMs/edges. It is observed from {Fig}.~4, the proposed template search algorithm offers an average running time of 0.013~seconds for searching one template, rather than averagely 0.298~seconds of ESA, in large problem size cases.

To investigate the factors that influence the running time of our proposed algorithm, we focus on two UAVs with the same signal coverage as service requestors, each carries a graph task (type 2). Based on which, {Fig}.~5 presents the performance comparisons of running time and templates count between three VCs with the same number of SPs and VMs, but different topologies (e.g., different number of edges). Specifically, each red dotted rectangle indicates a possible candidate of the component with ${\mathcal{D}}^c=5$ in graph task type 2. {Fig}.~5(a) and {Fig}.~5(b) focus on the VCs with the same number of SPs/VMs/edges (12/14/13), and the different topologies reflected by edge $4$ and edge $4'$ (the yellow edges). Apparently, the same amount of resources (SPs/VMs) and V2V connections (edges) can bring various number of possible candidates to components during mapping, which leads to significant differences of both the running time, and the templates count. Theoretically, the more candidates of a component will bring a larger searching space during mapping, and thus leads to higher
running time and more templates even under the same amount of resources (e.g., the same number of SPs and VMs). Comparatively, {Fig}.~5(c) shows a further complicated VC structure containing 12 SPs, 14 VMs and 22 edges, which enables more candidates of the component with ${\mathcal{D}}^c=5$ in graph task type 2. Correspondingly, obtaining all the templates of mapping the considered graph tasks to the VC in {Fig}.~5(c) results in a larger running time. Compared with the proposed algorithm, the running time of ESA in {Fig}.~5(a), {Fig}.~5(b) and {Fig}.~5(c) stay around 7000~seconds owing to that the computation complexity of which relies mainly on the total number of components, SPs and VMs.

\subsection{Performance Comparisons of The Value of $\mathcal{F}(\bm{x},\bm{q})$}
\begin{figure*}[t!]
\centering
\subfigure[]{\includegraphics[width=.322\linewidth]{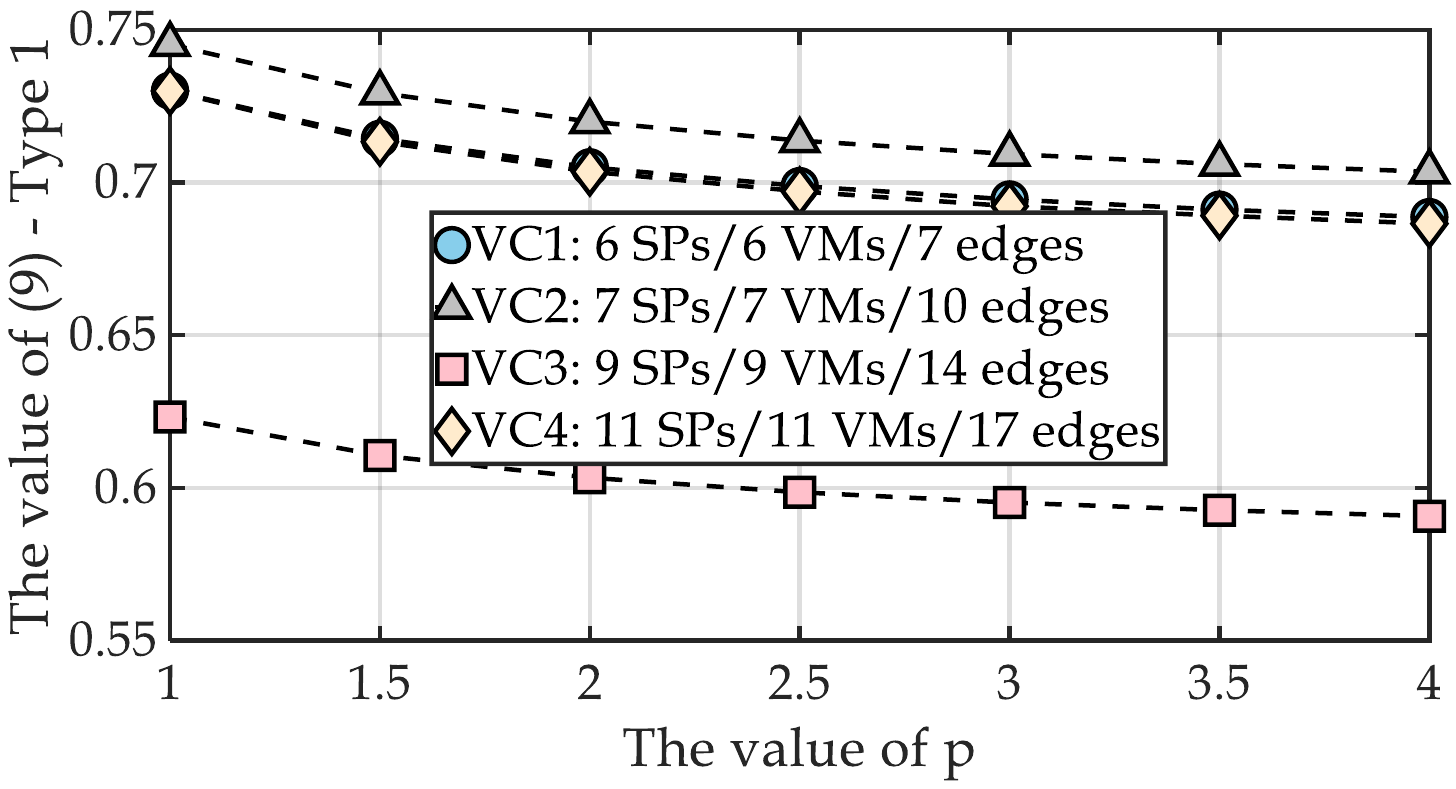}}
\subfigure[]{\includegraphics[width=.315\linewidth]{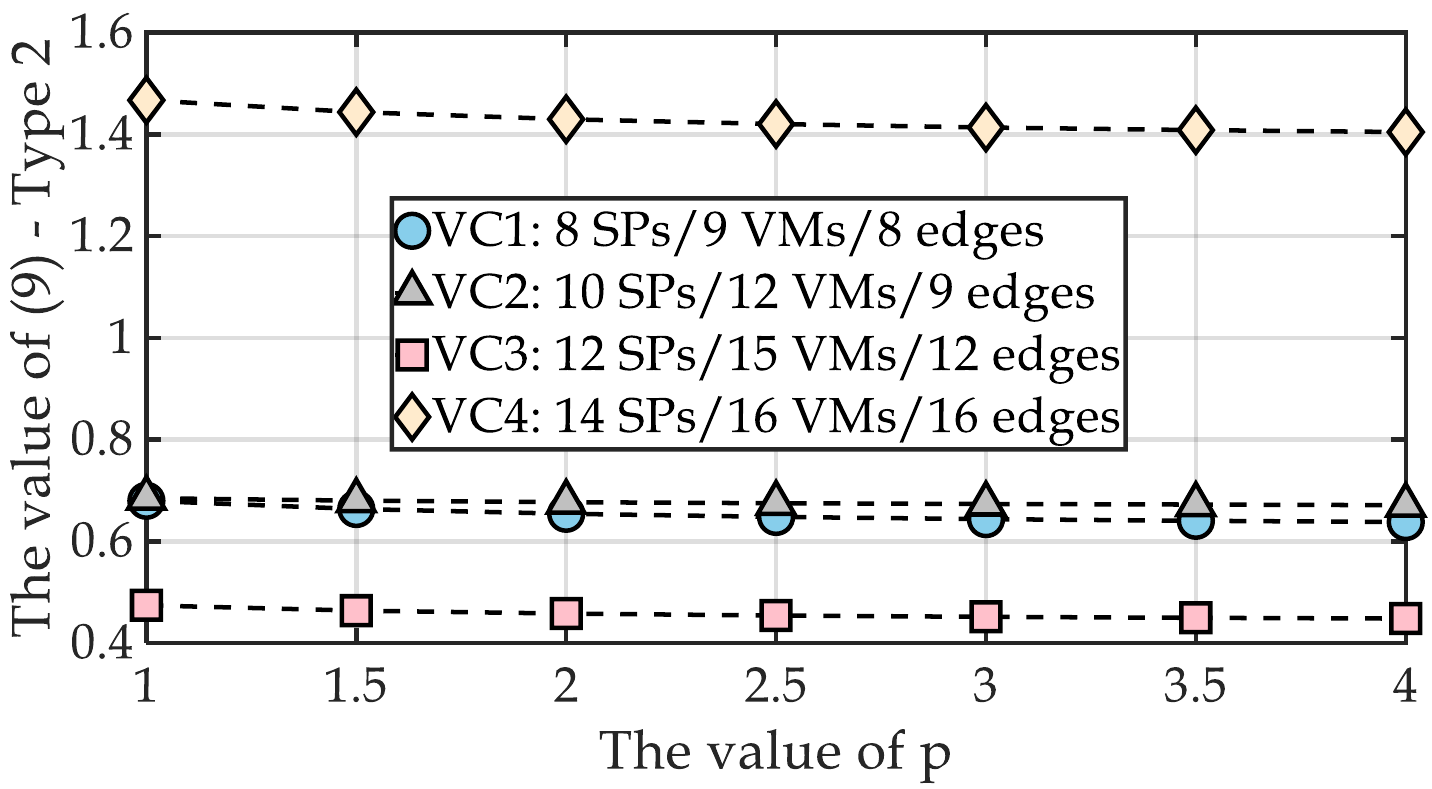}}
\subfigure[]{\includegraphics[width=.322\linewidth]{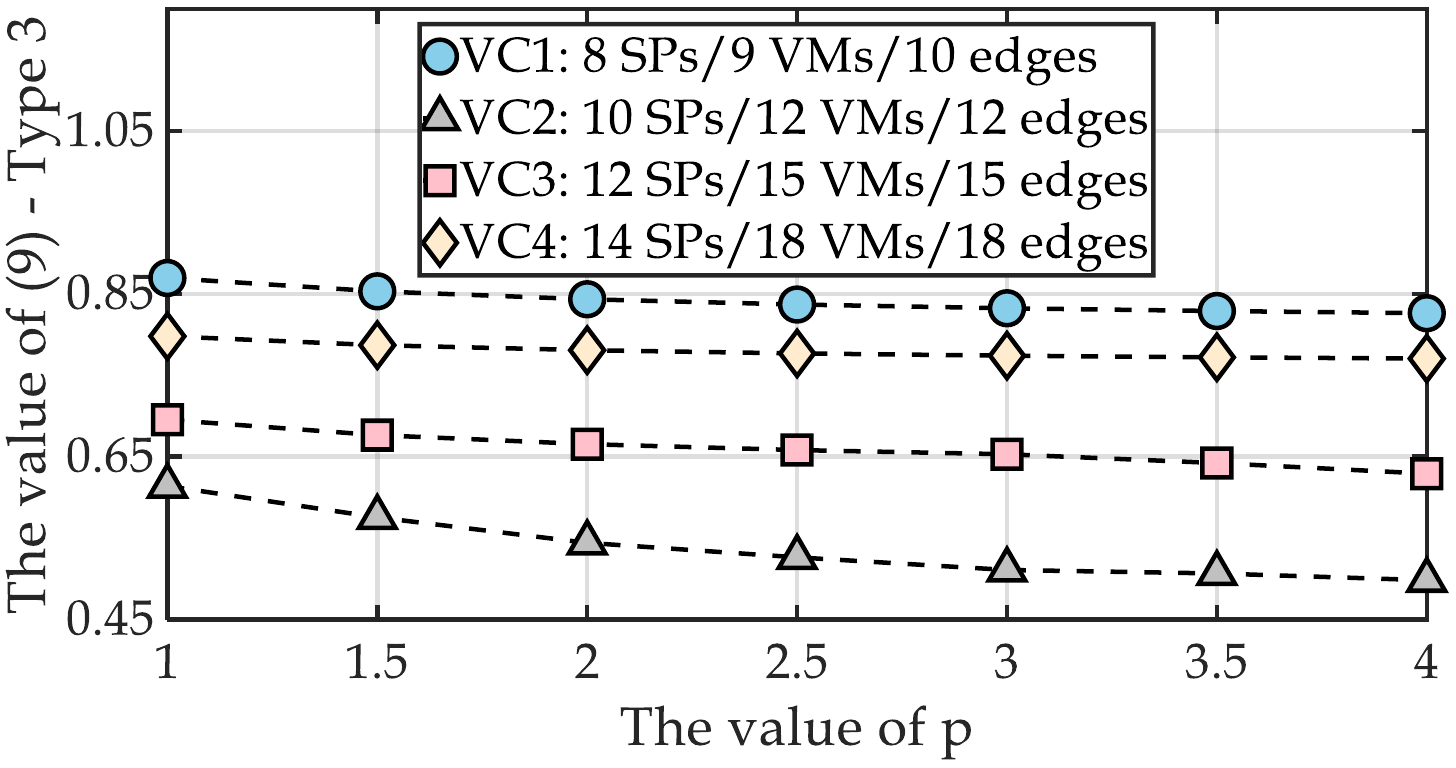}}
\caption{Performance comparison considering different values of $p$.}
\end{figure*}
\begin{figure}[t!]
\centering
\subfigure[]{\includegraphics[width=.48\linewidth]{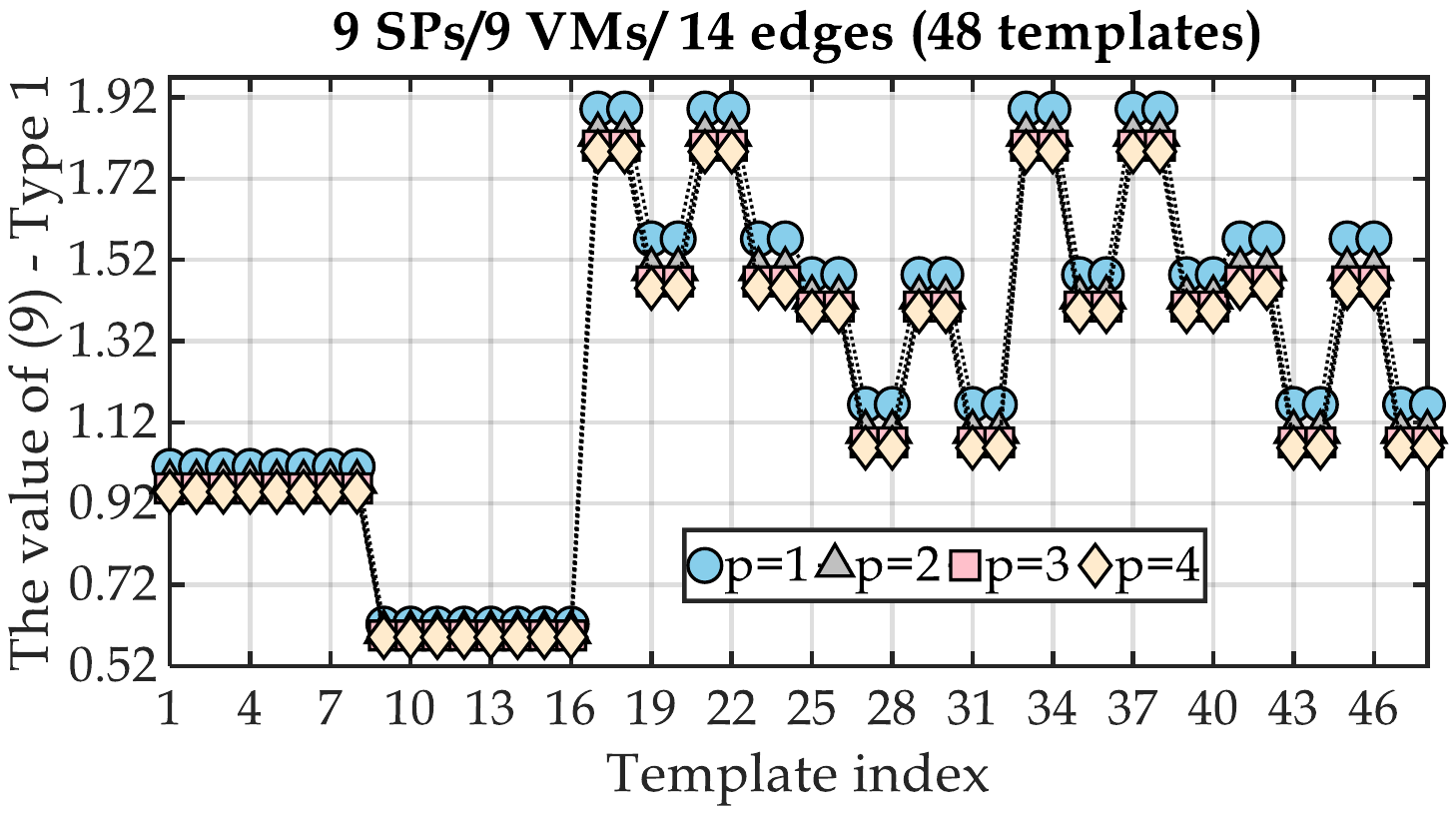}}
\subfigure[]{\includegraphics[width=.48\linewidth]{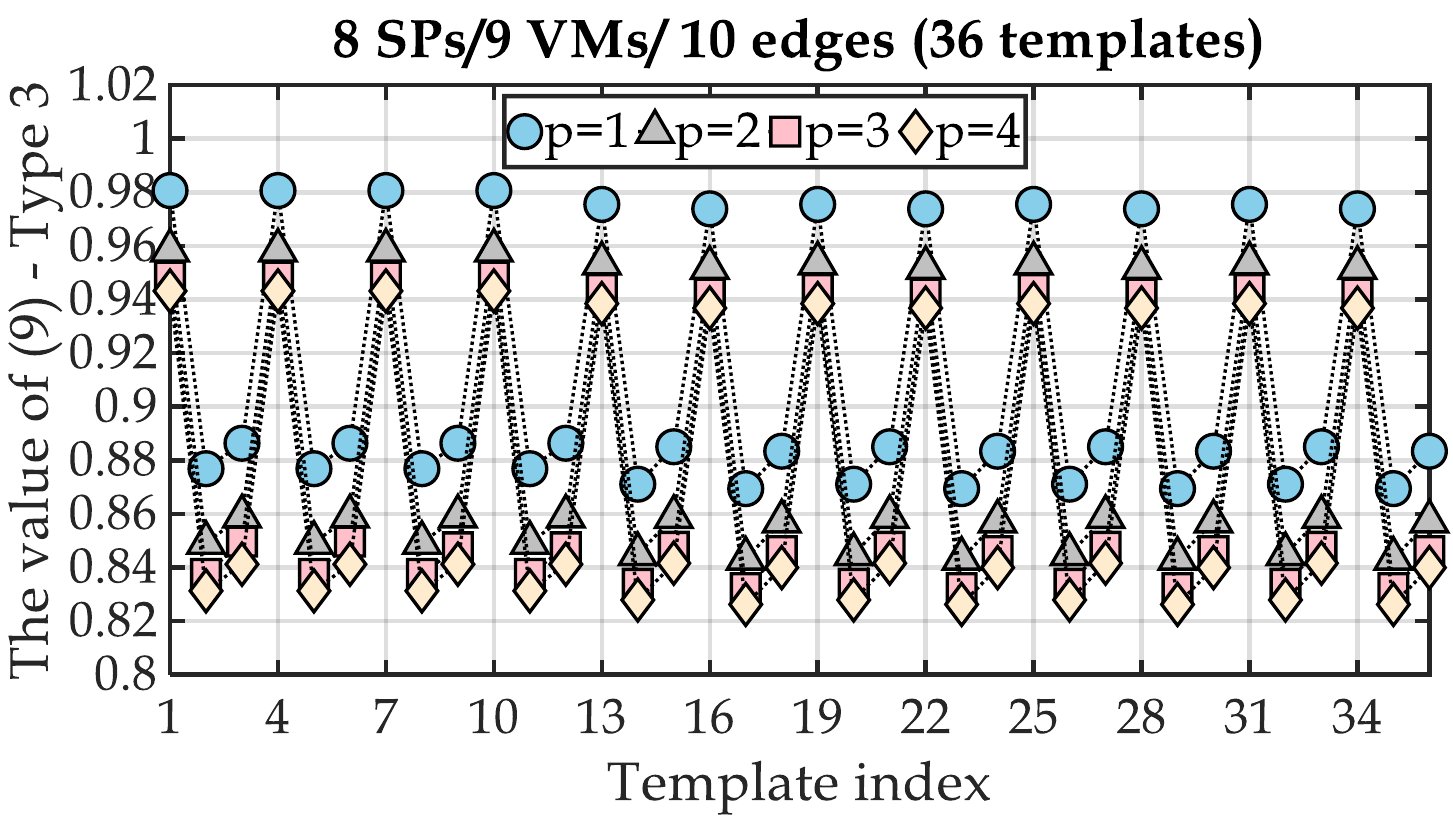}}
\caption{The relevant values of $\mathcal{F}(\bm{x},\bm{q})$ of each template considering different $p$: a). associated with VC3 in Fig. 7(a); b). associated with VC1 in Fig.7(c).}
\end{figure}
\noindent
Given the templates obtained from the proposed template search algorithm, Fig. 6 presents the performance comparison of the value of the objective function $\mathcal{F}(\bm{x},\bm{q})$ given in (9), between the proposed power allocation algorithm, and several baseline methods listed below.

\noindent
1) \textbf{Uniform allocation (UA)}: for each given template, the transmission power is uniformly allocated to each SP. The algorithm fails when cannot meet constraint (C13).

\noindent
2) \textbf{Random allocation (RA)}: for each given template, the transmission power is randomly allocated to each SP, until find a feasible allocation solution that meets constraint (C13).

\noindent
3) \textbf{Channel condition preferred allocation (CCPA)}: for each given template, a A2G channel with better condition (e.g, larger SNR) is allocated with more transmission power, while meeting constraint (C13).

\noindent
4) \textbf{Structure-preserved simulated annealing (SPSA)~\cite{18}}: for each given template, the transmission power is allocated to each SP via simulated annealing algorithm, while meeting constraint (C13).

In various small problem size cases where each considers one UAV and a couple of SPs, Fig. 6(a), Fig. 6(b) and Fig. 6(c) reveal that the performance of the value of $\mathcal{F}(\bm{x},\bm{q})$ greatly outperform the baseline methods UA, RA, CCPA, and SPSA, when applying the proposed power allocation algorithm with different values of $p$. Particularly, the cases where $p=3$ achieve better performance of the value of $\mathcal{F}(\bm{x},\bm{q})$ than those consider $p=1$, which commendably prove the theoretical idea of (13) (given in \textbf{Section 4.2}). Namely, a larger $p$ enables the value of a vector's $p$-norm to approach that of the $\infty$-norm. The values of $\mathcal{F}(\bm{x},\bm{q})$ of RA fluctuate irregularly owing to the randomness factor; while that of CCPA often stay at high values due to the deficiency of balancing various A2G channel conditions, which thus leads to larger data transmission rates and unsatisfactory task completion time. Furthermore, our proposed algorithm obtains far better performance than SPSA since the process of generating a new state in SPSA only considers discrete values of power, which pose difficulties searching for the whole solution space. The performance comparisons considering large problem size cases with more UAVs and SPs, as well as complicated VC structures are depicted in Fig. 6(d), Fig. 6(e) and Fig. 6(f). Similarly, our proposed algorithm can approach better performance than the baseline methods under both situations when $p=1$ and $p=3$.

\subsection{Performance Evaluation upon Considering Different Values of $p$}

\noindent
As mentioned in \textbf{Section 4.2}, the larger value of $p$ will bring a better performance on power allocation. Concretely, since a vector's $\infty$-norm describes the largest value (peak value) in this vector, while a vector's $p$-norm can approach the relevant $\infty$-norm upon increasing the value of $p$. Correspondingly, this section depicts the impact on 
$\mathcal{F}(\bm{x},\bm{q})$ when considering various $p$. Note that the proposed power allocation algorithm works indistinguishably and independently among different templates and UAVs, we focus on single UAV scenarios as examples. Fig. 7 demonstrates that for different graph task types and VC structures, upon increasing the value of $p$ can always bring a better solution for problem 
$\bm{\mathcal{P}_2}$ given in (12), and thus achieve a satisfying performance of the objective function $\mathcal{F}(\bm{x},\bm{q})$. 

Fig. 8 shows two examples of the changing process on the value of $\mathcal{F}(\bm{x},\bm{q})$, upon having various $p$ and different templates. Fig. 8(a) is associated with graph task type 1 and VC3 of Fig. 7(a), where 48 templates are obtained from applying the proposed template search algorithm. Apparently, under each given template, a larger $p$ achieves a lower $\mathcal{F}(\bm{x},\bm{q})$, and the best graph task and power allocatipon solution can be obtained by comparing through all the templates. Similar conclusion can be found from Fig. 8(b) under 36 templates, which is related to graph task type 3 and VC1 shown in Fig. 7(c).

\section{Conclusion}

\noindent 
This paper studies the energy-aware graph task scheduling problem in SD-AGV networks. To achieve the trade-off upon minimizing the task completion time and energy consumption, as well as the data exchange cost, the problem is formulated as a MINLP problem which is NP-hard. An efficient decoupled approach is proposed by separating the template search stage from the power allocation stage. In the former stage, an effective algorithm is presented to search for all the subgraph isomorphisms between the graph tasks and the VC structure. For the latter stage, we introduce an applicable power allocation optimization algorithm by applying convex optimization techniques. The effectiveness of the proposed approach is revealed through comprehensive simulations. Several future research directions could involve improving the computation efficiency of the template search algorithm, and considering the optimization of the UAV's path trajectory to achieve better task scheduling performance.



\ifCLASSOPTIONcaptionsoff
 \newpage
\fi

\vfill

\end{document}